%% file: main.tex
\documentclass[11pt]{article}

\usepackage[a4paper,margin=1in]{geometry}

\usepackage{graphicx}
\usepackage{dcolumn}
\usepackage{bm}
\usepackage[numbers,square]{natbib}
\usepackage{upgreek}
\usepackage{makecell}
\usepackage{xcolor}
\usepackage{amssymb} 
\usepackage{amsmath}
\usepackage{tikz}
\usepackage{subfigure}
\usepackage{caption}
\usepackage{booktabs}
\newcommand{\symo}[1]{\tikz\fill[#1] (0,0) circle (3pt);}

\newcommand{\syms}[1]{\tikz\fill[#1] (-3pt,-3pt) rectangle (3pt,3pt);}

\newcommand{\symD}[1]{%
\tikz[baseline=-0.6ex]
\fill[#1]
(0,5pt) -- (3.5pt,0) -- (0,-5pt) -- (-3.5pt,0) -- cycle;
}

\newcommand{\symu}[1]{\tikz\fill[#1] (0,3.5pt) -- (3.5pt,-3.5pt) -- (-3.5pt,-3.5pt) -- cycle;}

\newcommand{\symr}[1]{\tikz\fill[#1] (-3.5pt,3.5pt) -- (3.5pt,0) -- (-3.5pt,-3.5pt) -- cycle;}

\newcommand{\symv}[1]{\tikz\fill[#1] (-3.5pt,3.5pt) -- (3.5pt,3.5pt) -- (0,-3.5pt) -- cycle;}

\newcommand{\syml}[1]{%
\tikz[baseline=-0.6ex]
\fill[#1] (3.5pt,3.5pt) -- (-3.5pt,0) -- (3.5pt,-3.5pt) -- cycle;
}

\newcommand{\symp}[1]{%
\tikz[baseline=-0.6ex]
\fill[#1]
(90:4.5pt) -- (234:4.5pt) -- (18:4.5pt) -- (162:4.5pt) -- (306:4.5pt) -- cycle;
}

\newcommand{\symh}[1]{%
\tikz[baseline=-0.6ex]
\fill[#1]
(90:4.5pt) -- (210:4.5pt) -- (330:4.5pt) -- cycle
(270:4.5pt) -- (30:4.5pt) -- (150:4.5pt) -- cycle;
}

\newcommand{\symstar}[1]{%
\tikz[baseline=-0.6ex]
\draw[#1,line width=0.9pt,line cap=round]
  (-4pt,0) -- (4pt,0)      
  (0,-4pt) -- (0,4pt)      
  (-2.83pt,-2.83pt) -- (2.83pt,2.83pt)   
  (-2.83pt,2.83pt) -- (2.83pt,-2.83pt);  
}

\newcommand{\symx}[1]{%
\tikz[baseline=-0.6ex]
\draw[#1,line width=1.5pt,line cap=round]
  (-3.5pt,-3.5pt) -- (3.5pt,3.5pt)
  (-3.5pt,3.5pt) -- (3.5pt,-3.5pt);
}

\newcommand{\symplus}[1]{%
\tikz[baseline=-0.6ex]
\draw[#1,line width=1.5pt,line cap=round]
  (-4.5pt,0) -- (4.5pt,0)
  (0,-4.5pt) -- (0,4.5pt);
}

\definecolor{c1}{rgb}{0,0.447,0.741}
\definecolor{c2}{rgb}{0.850,0.325,0.098}
\definecolor{c3}{rgb}{0.929,0.694,0.125}
\definecolor{c4}{rgb}{0.494,0.184,0.556}
\definecolor{c5}{rgb}{0.466,0.674,0.188}
\definecolor{c6}{rgb}{0.301,0.745,0.933}

\usepackage{titlesec}

\titleformat{\section}
    {\normalfont\large\bfseries}
    {\thesection.}
    {0.5em}
    {}

\titleformat{\subsection}
    {\normalfont\normalsize\bfseries}
    {\thesubsection}
    {0.5em}
    {}

\titleformat{\subsubsection}
    {\normalfont\normalsize\itshape}
    {\thesubsubsection}
    {0.5em}
    {}

\titlespacing*{\section}
    {0pt}{12pt}{6pt}

\titlespacing*{\subsection}
    {0pt}{10pt}{4pt}

\titlespacing*{\subsubsection}
    {0pt}{8pt}{3pt}

\newenvironment{myabstract}
{
    \par
    \noindent
    \rule{\textwidth}{0.5pt}

    \vspace{6pt}

    \begin{list}{}{
        \setlength{\leftmargin}{15pt}
        \setlength{\rightmargin}{15pt}
        \setlength{\listparindent}{0pt}
        \setlength{\itemindent}{0pt}
    }

    \item[]

    \textbf{Abstract}

    \vspace{3pt}

    \small
}
{
    \normalsize

    \end{list}

    \vspace{3pt}

    \noindent
    \rule{\textwidth}{0.5pt}

    \vspace{10pt}
}

\title{
    \vspace{-20mm}
    \textbf{
        Coupled Analytical Model for the Internal Boundary Layer Height, Wall Shear Stress and Mean Velocity Behind a Surface Roughness Transition in Boundary-Layer Flow
    }
}

\author{
    Kingshuk Mondal$^{1}$,
    Niranjan S. Ghaisas$^{1,2,3}$\\[5pt]
    \small
    $^{1}$Department of Mechanical and Aerospace Engineering\\
    \small
    Indian Institute of Technology Hyderabad, Telangana, 502285, India\\
    \small
    $^{2}$Department of Climate Change \\
    \small
    Indian Institute of Technology Hyderabad, Telangana, 502285, India\\
    \small
    $^{3}$Greenko School of Sustainability\\
    \small
    Indian Institute of Technology Hyderabad, Telangana, 502285, India
}

\date{}

\begin{document}

\maketitle


\begin{myabstract}

A coupled analytical framework is developed to predict the internal boundary layer (IBL) height, wall shear stress and the mean velocity profile downstream of a surface roughness transition in a neutral atmospheric boundary-layer. The new model combines a three-layer analytical velocity formulation with a modified diffusion analogy for the IBL growth rate. The diffusion analogy links the growth rate of the IBL to turbulent diffusion and to mean vertical advection. Our formulation rectifies physical inconsistencies in a previous model that assumed that the turbulent diffusion is controlled only by the upstream surface conditions and that the characteristic streamwise velocity difference caused by streamline displacement is independent of downstream distance from the roughness transition. Key model parameters, namely the turbulent diffusion coefficient and the eddy viscosity augmentation coefficient, are modelled as functions of the upstream-to-downstream aerodynamic roughness length ratio. The coupled model, along with three other analytical models, is tested against twelve datasets covering a wide range of roughness ratios, including wind-tunnel experiments and large-eddy simulations. The new model accurately predicts wall shear stress, mean velocity, and IBL height for all combinations of upstream and downstream roughness values evaluated, for both smooth-to-rough and rough-to-smooth transitions.

\vspace{4pt}


\end{myabstract}

\section{Introduction}
\label{sec:intro}
\input{files/introduction2}

\section{LES methodology and simulations}
\label{sec:les}

\subsection{LES methodology}
\label{ssec:les}
\input{files/les_methodology}

\subsection{Simulation details and datasets}
\label{ssec:cases}
\input{files/cases}

\section{Analytical modelling framework}
\label{sec:model}
\subsection{Modelling the streamwise velocity and wall shear stress}
\label{ssec:u_model}
\input{files/blm2020}

\subsection{Modelling the internal boundary layer height}
\label{ssec:ibl_model}
\input{files/ibl_model}


\section{Results and discussion}
\label{sec:results}
\input{files/results}

\section{Conclusion}
\label{sec:conclusion}
\input{files/conclusion}

\section*{Acknowledgments}

KM acknowledges support from the Prime Minister's Research Fellowship (PMRF). NSG thanks SERB/ANRF for enabling this work under the Core Research Grant (No. CRG/2022/006735) and Advanced Research Grant (No. ARG/2025/010314/ENS).

\appendix

\section{Derivation of the streamwise velocity model}


\label{app:utrans}
\input{files/app1}


\bibliographystyle{unsrtnat}

\bibliography{references}

\end{document}

%% file: files/introduction2.tex
The atmospheric boundary layer (ABL) flow is highly turbulent compared to the free atmosphere \cite{wyngaard2010turbulence} and is affected by shear, buoyancy, surface forcing, and Coriolis effects \cite{cheng2021prf,liu2021prl}. Since the ABL is the lower-most part of the atmosphere, directly adjacent to the Earth's surface, the flow in the ABL is directly affected by surface conditions. During short time periods around sunset and sunrise, thermal gradients subside and the effects of buoyancy are not significant \cite{stull1988BL}. Moreover, for flow developing over short horizontal scales, the Rossby number is large and the effects of the geostrophic forces are not felt near the ground \cite{wyngaard2010turbulence}. As a result, a common idealization in ABL studies is to neglect the effects of thermal stratification and Coriolis forces, which leads to a truly neutral ABL \citep{zilitinkevich2002integral,joshi2022surface, klemmer2024prf}.   
The structure of a truly neutral ABL is governed primarily by the mechanical shear imposed by the underlying surface. The surface can be considered as a fully-rough wall, where the typical height of the surface irregularities, $k$, is much larger than the viscous sub-layer but much smaller than the boundary layer height, $\delta$ \cite{jimenez2004}. For such an ABL flow with the ratio $\delta/k\gg1$, the detailed geometry of the individual roughness elements does not matter \cite{jimenez2004}. The drag imposed by the roughness elements can then be characterized by an aerodynamic roughness length, $z_0$, which can be related to $k$ through simple empirical relations \cite{jimenez2004}.

\begin{figure}
    \centering
    \includegraphics[width=0.8\linewidth]{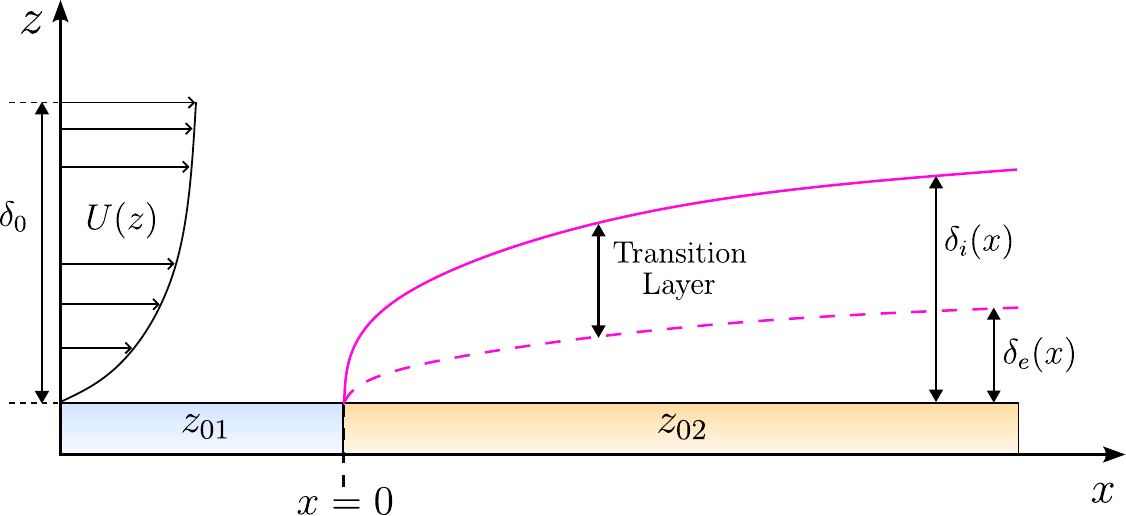}
    \caption{Schematic of the boundary-layer flow over an abrupt roughness transition from $z_{01}$ to $z_{02}$ at $x=0$. $\delta_0$ is the upstream boundary layer height. The solid magenta line represents the internal boundary layer (IBL; height $\delta_i$) and the dashed magenta line represents the equilibrium boundary layer (EBL; height $\delta_e$)}
    \label{fig:schematic}
\end{figure}

Heterogeneity in land-surface properties arises from transitions between different surface types, such as sea-to-land or agricultural fields to forests. This change in surface type can be characterized by a change in the value of $z_0$. Figure~\ref{fig:schematic} illustrates a canonical form of surface heterogeneity where the roughness length abruptly changes from $z_{01}$ to $z_{02}$ in the direction perpendicular to that of the mean flow. Here, $x$ and $z$ are the streamwise and vertical directions, respectively. The transition is referred to as smooth-to-rough (S$\rightarrow$R) when $z_{01}<z_{02}$ and rough-to-smooth (R$\rightarrow$S) when $z_{01}>z_{02}$. The incoming flow is assumed to be in equilibrium with the upstream roughness $z_{01}$ and has a boundary layer height of $\delta_0$. As the incoming flow encounters the roughness transition at $x=0$, an internal boundary layer (IBL) is formed of height $\delta_i(x)$, which acts as a transition between the momentum fluxes imposed by the upstream and downstream surfaces. Below the height $\delta_e$, the flow is fully adjusted to the new surface conditions, and this region is called the equilibrium boundary layer (EBL). Within the IBL, the wall shear stress and turbulence quantities are modified by the new surface conditions, while above $\delta_i$ these quantities retain the influence of the upstream surface. 

The flow over an abrupt roughness transition has been extensively studied over the years by wind tunnel experiments \cite{chamorro2009velocity,efros2011s2r,li2021r2s}, analytical models \cite{elliott1958growth,panofsky1964change,plate1967laboratory,taylor1969wind,chamorro2009velocity,ghaisasblm2020,li2022modelling} and numerical simulations \cite{bou2004large,abkar2012new,rouhi2019direct,anderson2020turbulent,mondal2023r2s,mondal2025s2r,cooke2025iblprf}. The experiments and simulations, including high-fidelity large eddy simulations (LES) have provided valuable insights and data that have been used to develop and refine analytical models. Several analytical models for the mean velocity \cite{panofsky1964change,chamorro2009velocity,ghaisasblm2020}, the turbulence intensity \cite{mondal2023r2s, mondal2025s2r} and for the wall shear stress \cite{abkar2012new, ghaisasblm2020} have been proposed. All these models require the IBL height as an input. A number of analytical and empirical models for $\delta_i$ have been proposed in the literature. The earliest such model was proposed by \citet{elliott1958growth}, who suggested that the IBL height grows as a power law, $\delta_i \sim x^{0.8}$. Subsequent works by \citet{wood1982internal}, \citet{pendergrass1984dispersion}, \citet{jegede1999study} and \citet{cheng2002near} reported similar observations, although the proportionality functions were different in each study. \citet{miyake1965} linked the IBL growth to the vertical diffusion of a plume, specifically assuming it to be proportional to the vertical diffusion intensity. \citet{panofskyDutton1984} hypothesized that the vertical diffusion intensity is proportional to the downstream friction velocity and obtained an implicit relation for the IBL height by assuming a logarithmic-law relation for the velocity profile downstream of the roughness transition.  
\citet{savelyev2001notes} modelled the proportionality constant of \citeauthor{miyake1965}'s model as a function of $\ln{(z_{02}/z_{01})}$ by analysing a large number of experimental datasets. In a later work, \citet{savelyev2005internal} (henceforth referred to as `ST-model') modified \citeauthor{miyake1965}'s model by incorporating the mean wall-normal velocity to account for the displacement of streamlines. The model was specified with up to two parameters, the values of which were determined by fitting the model to several experimental datasets. The above-mentioned models were not comprehensive, since they either predicted the wall shear stress assuming the IBL height, or assumed a velocity profile (rather than predicting it) to derive predictive expressions for the other quantities of interest. 

The modelling framework of \citet{li2022modelling} is among the only comprehensive models that predict the mean velocity, wall shear stress as well as IBL height.  \citet{li2022modelling} built upon \citeauthor{elliott1958growth}'s \cite{elliott1958growth} momentum integral model and extended it to finite-thickness boundary layer flows. This model modified the velocity profile downstream of the roughness transition by augmenting the wake function \cite{jones2001wake} to the logarithmic law of the wall. The shear stress was modelled as a non-linear function of the ratio of the vertical height ($z$) and the outer boundary layer height ($\delta_c$). The modified velocity profile and the shear stress were substituted into the momentum equation and solved numerically to compute $u_{*1}$, $u_{*2}$, $\delta_i$ and $\delta_c$ downstream of the roughness transition. 

Recently, \citet{ding2025ibl} extended the ST-model to predict IBL growth in stably stratified boundary layers. They proposed empirical relations for $\sigma_w$ and the mean streamwise velocity profiles, which incorporate an explicit dependence on the Obukhov length ($L_0$) to make the model applicable to stably stratified flows. The framework of \citet{ding2025ibl} relied on empirically fitted parameters to model the mean velocity and shear stress profiles. The empirical relations were derived by studying several experiments pertaining only to S$\rightarrow$R transitions \cite{ding2023exp}.  

Several of the aforementioned empirical and semi-empirical models for the IBL height \cite{elliott1958growth,wood1982internal,pendergrass1984dispersion,panofskyDutton1984,jegede1999study,cheng2002near} were evaluated against LES data in our previous work \cite{mondal2023r2s,mondal2025s2r}. It was concluded that that no single IBL model is able to accurately predict the IBL height for both (S$\rightarrow$R and R$\rightarrow$S) roughness transitions with different roughness ratios, $m=z_{01}/z_{02}$. This highlights the need for developing a model for the IBL height, which is an important component in applications such as wind power meteorology and air pollution meteorology. Furthermore, the discussion above highlights the need to develop a framework that predicts the velcoity profile, wall shear stress as well as IBL height behind a roughness transition in a comprehensive and physics-based manner.

The purpose of this paper is to develop a comprehensive analytical modelling framework that simultaneously predicts the wall shear stress, mean velocity, and IBL height behind a surface roughness transition. This is achieved by modifying some of the key assumptions in the ST-model and coupling them with the analytical framework of \citet{ghaisasblm2020}. The coupled model is evaluated against twelve datasets covering a wide range of upstream surface roughness and upstream-to-downstream roughness ratio values. Six of these datasets are used to gain physical insights, justify the modelling assumptions and to calibrate model coefficients. The remaining six datasets are entirely unseen by the model and comparisons with these enable a comprehensive evaluation of the general applicability of our modelling framework. Our proposed model is shown to work well for S$\rightarrow$R as well as R$\rightarrow$S transitions. 

Section~\ref{ssec:les} briefly describes the LES code used to generate LES datasets and the cases simulated. The LES code has been well-validated previously with multiple experimental datasets, which is briefly reviewed in Section~\ref{ssec:cases}. Section~\ref{sec:model} describes the development of the analytical framework, followed by its evaluation in Section~\ref{sec:results}. Conclusions are presented in section~\ref{sec:conclusion}.

%% file: files/les_methodology.tex
The incompressible LES-filtered continuity and Navier-Stokes equations are solved to simulate the flow over a roughness transition using a modified concurrent-precursor simulation technique \cite{stevens2014concurrent,mondal2025s2r}. The three coordinate directions are $x_i$ with $i=1,2,3$, or equivalently, $x, y$ and $z$. The filtered variables that are solved for are the three components of velocity, $u_i$, and pressure, $p$. 

\begin{figure}[h!]
    \centering
    \subfigure[]{\includegraphics[width=0.65\linewidth] {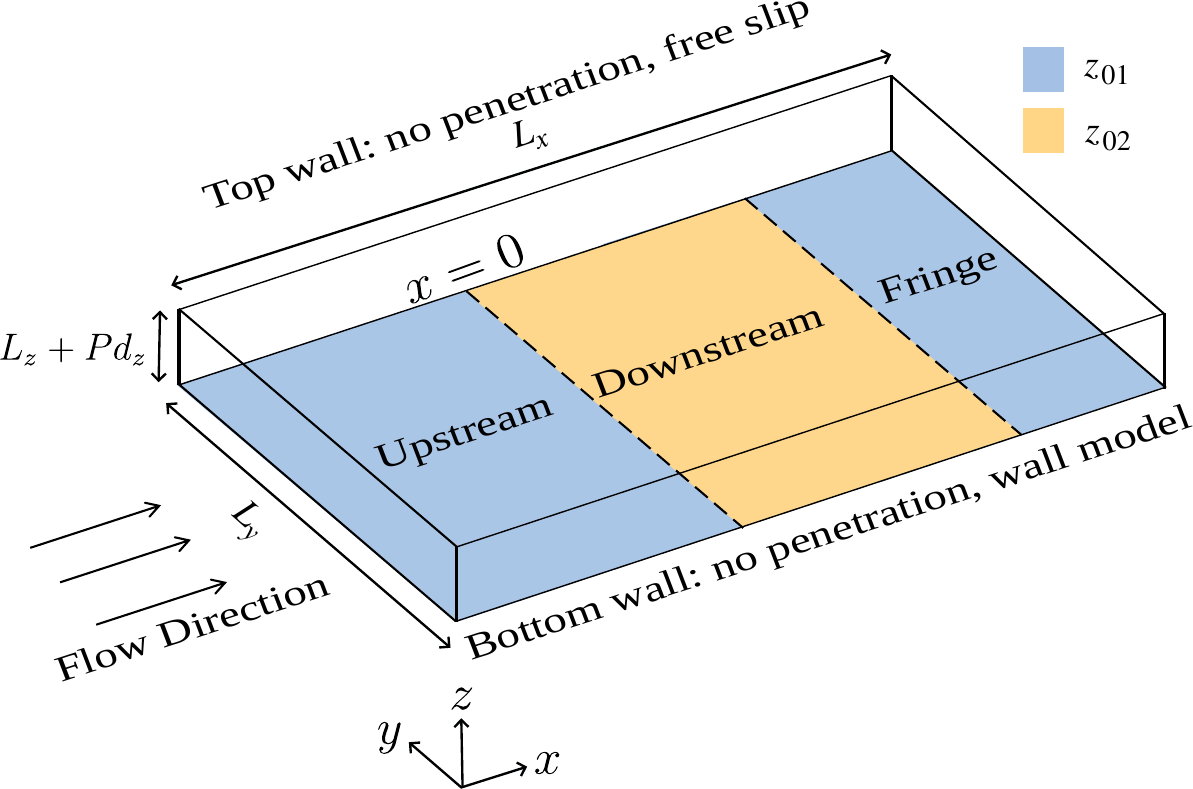} \label{fig:les_schematic}}
    \subfigure[]{\includegraphics[width=1.1\linewidth]{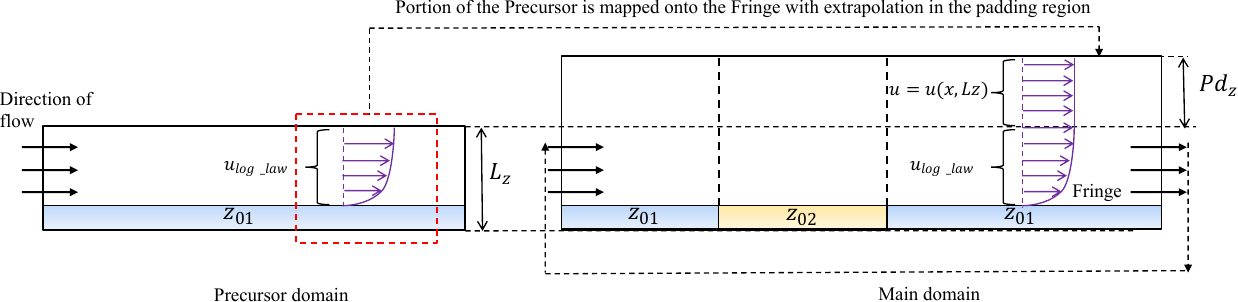} \label{fig:padding}}
    \caption{Schematic (not to scale) of (a) the main domain showing different regions and the top and bottom boundary conditions. $z_{01}$ and $z_{02}$ are the aerodynamic roughness lengths upstream and downstream of the roughness transition at $x=0$. The fringe region has the roughness $z_{01}$, and (b) the precursor and main domains. The main domain is padded by height $Pd_z$ compared to the height of the precursor domain, $L_z$. The target velocity for the fringe in the padded region is that at the height $L_z$. The flow at the exit of the fringe acts as an input to the main domain due to the periodic boundary conditions in the streamwise direction}
    \label{fig:schematics}
\end{figure}

Two computational domains, `precursor' and `main', are used with the same horizontal ($L_x$ and $L_y$) dimensions, with $x$ and $y$ being the streamwise and spanwise directions, respectively. The vertical dimensions are $L_z$ and ($L_z+Pd_z$), respectively, of the precursor and main domains. Here, $z$ is the vertical coordinate and $Pd_z$ is the height of the additional padding in the main domain, which will be discussed later in this section. The precursor domain has a homogeneous surface with aerodynamic roughness length $z_{01}$ and an imposed, constant, pressure gradient equal to $-u_{*1}^2/L_z$, where $u_{*1}$ is the friction velocity. The main domain (shown in Fig.~\ref{fig:schematics}(a)) has an upstream portion with roughness $z_{01}$ followed by a transition (at $x=0$) to a surface with roughness $z_{02}$. The last portion of the main domain is a `fringe' region which has roughness $z_{01}$, i.e. same as the precursor domain and the upstream portion of the main domain. 

In the fringe region, the flow is nudged towards the same conditions as in the precursor domain using an additional forcing term that, for the $i-$direction momentum equation, is proportional to $-(u_i-u_i^{targ})$. Here, $u_i^{targ}$ is the `target' velocity field, which is prescribed as detailed in \citet{mondal2025s2r} and consists of two parts. Up to the height $L_z$, the target velocity is the same as that in the precursor domain. Above the height $L_z$, the target velocity values at $z=L_z$ are extended into the padded region. The vertical padding of height $Pd_z$ is provided in the main domain (as seen in Fig.~\ref{fig:schematics}(b)) to ensure that insufficient domain height does not artificially impede the development of the flow downstream of the roughness transition. Here, we use $Pd_z=L_z$ in accordance with our previous study \cite{mondal2025s2r}. 

Periodic boundary conditions are imposed in the streamwise and spanwise directions. The forcing term in the fringe region coupled with the streamwise periodicity of the main domain ensures that the fully-developed boundary-layer flow from the precursor is fed as an input to the main domain upstream of the surface roughness transition at $x=0$. Free-slip conditions ($\partial u/\partial z = \partial v/\partial z = 0$) are applied on the top boundary. The shear stress at the bottom wall is prescribed through the algebraic wall model proposed by \citet{bou2004large} that has been shown to be appropriate for flow over heterogeneous surfaces \cite{mondal2023r2s, mondal2025s2r}. No-penetration  ($w=0$) conditions are imposed on both the top and bottom boundaries.

Our in-house `PadeOps-igrid' code \cite{padeops} is used for the simulations. This code has been developed over the years to perform LES of atmospheric flows \cite{ghate2017abl,mondal2023r2s,mondal2025s2r} and of the flow over wind turbines \cite{ghaisas2020effect,howland2020optimal,kethavath2022large,kethavath2024effect,kethavath2025effect}. The code uses Fourier-pseudo-spectral discretization in the streamwise and spanwise directions and a 6\textsuperscript{th}-order staggered compact finite-difference scheme in the vertical direction. The anisotropic minimum dissipation model \cite{rozema2015amd,abkar2016amd} is employed subgrid-scale closure. The Reynolds number (computed using the free-stream velocity, the height of the boundary layer and the viscosity of air) exceeds $10^7$ for typical atmospheric flows. As a result, the viscous sublayer, where the direct effect of viscosity is significant, is confined to an extremely thin layer near the surface and cannot be resolved on moderately sized grids. For computational efficiency, we employ the standard practice \cite{calaf2010large, stevens2014concurrent, anderson2020turbulent} of dropping the viscous terms from the Navier-Stokes equations and introducing an additional shear stress at the bottom of the wall using a wall model \cite{bou2004large}.

%% file: files/cases.tex
\begin{table}[htbp]
\centering
\caption{List of LES simulations (Set A) used to develop the analytical model. $z_{01}$ and $z_{02}$ are roughness values upstream and downstream of the transition, respectively, and their ratio is $m=z_{01}/z_{02}$. The LES cases that have been reported in our previous work have a citation under `Label'}
\label{tab:cases_train}
\makebox[\textwidth][c]{
\begin{tabular}{ccccccc}
\toprule
No. & Symbol & Label & $z_{01}/\delta_0$  & $m$   &   Transition Type &  Reference Experiment\\
\midrule
1 & \symo{c1!80} & EK-SR \cite{mondal2025s2r} & $3.0\times 10^{-5}$ & 1/150 & S$\rightarrow$R &  \citet{efros2011s2r}  \\
2 & \syms{c2!80} & GG-SR \cite{mondal2025s2r} & $2.5\times 10^{-4}$ & 1/4.34 & S$\rightarrow$R &  \citet{gul2022s2r}\\
3 & \symD{c3!80} & LI-RS1 & $4.6\times 10^{-5}$ & 15 & R$\rightarrow$S &  \citet{li2021r2s} \\
4 & \symu{c4!80} & LI-RS2 \cite{mondal2023r2s} & $2.0\times 10^{-5}$ & 21 & R$\rightarrow$S &  \citet{li2021r2s} \\
5 & \symr{c5!80} & KM-RS1 \cite{mondal2023r2s} & $1.2\times 10^{-3}$ & 83.3 & R$\rightarrow$S & \citet{chamorro2009velocity}  \\
6 & \symv{c6!80} & KM-RS2 \cite{mondal2023r2s} & $1.2\times 10^{-3}$ & 125 & R$\rightarrow$S & -- \\
\bottomrule
\end{tabular}%
}
\end{table}

Twelve LES are reported in this study, divided into two sets. Set A (Table~\ref{tab:cases_train}) is used for gaining insight and for development of the analytical model, while Set B (Table~\ref{tab:cases_test}) is used to evaluate the analytical model. These twelve LES consist of a mix of R$\rightarrow$S and S$\rightarrow$R transitions, different upstream roughness lengths ($z_{01}/\delta_0$) and different roughness ratios ($m=z_{01}/z_{02}$). 
Wind-tunnel experimental data has been reported previously in the literature corresponding to Cases 1--5 and 11 as noted in Tables~\ref{tab:cases_train} and \ref{tab:cases_test}. Since our focus is on geophysical flows, all experimental datasets are chosen such that $\delta/k$ is at least 80 \cite{jimenez2004}.  
Our LES framework has been validated rigorously with these experimental data, including two S$\rightarrow$R transition (Cases 1, 2) and two R$\rightarrow$S transition (Cases 4, 5), as reported in our previous work \cite{mondal2023r2s,mondal2025s2r}. New simulations are carried out for Cases 3, 9, 11 and 12.  
The domain sizes and number of grid points used for these simulations are listed in Table~\ref{tab:domain_size}. We arrive at these following sensitivity studies reported in \cite{mondal2023r2s, mondal2025s2r} and the reported sizes are sufficient to get statistically converged results. It is worth noting that the streamwise domain size and number of streamwise grid points for the S$\rightarrow$R cases are larger than for the R$\rightarrow$S cases because experimental data for the former are typically available till farther downstream locations.

\begin{table}[htbp]
\centering
\caption{List of LES simulations (Set B) used to evaluate the analytical model. $z_{01}$ and $z_{02}$ are roughness values upstream and downstream of the transition, respectively, and their ratio is $m=z_{01}/z_{02}$. The LES cases that have been reported in our previous work have a citation under `Label'}
\label{tab:cases_test}
\begin{tabular}{ccccccc}
\toprule
No. & Symbol & Case Label & $z_{01}/\delta_0$  & $m$   &   Transition Type &  Reference Experiment\\
\midrule
7 & \symstar{c6!80} & KM-SR1  \cite{mondal2025s2r} & $3.0\times 10^{-5}$ & 1/113 & S$\rightarrow$R &  -- \\
8 & \symplus{c4!80} & KM-SR2  \cite{mondal2025s2r} & $3.0\times 10^{-5}$ & 1/77 & S$\rightarrow$R & -- \\
9 & \symh{c3!80} & KM-SR3 & $3.0\times 10^{-5}$ & 1/60 & S$\rightarrow$R & -- \\
10 & \symx{c5!80} & KM-SR4  \cite{mondal2025s2r} & $3.0\times 10^{-5}$ & 1/40 & S$\rightarrow$R &  -- \\
11 & \syml{c1!80} & LI-RS3  & $2.0\times 10^{-5}$ & 21.75 & R$\rightarrow$S &  \citet{li2021r2s}\\
12 & \symp{c2!80} & KM-RS3 & $3.0\times 10^{-5}$ & 40 & R$\rightarrow$S &   -- \\
\bottomrule
\end{tabular}
\end{table}

\begin{table}[htbp]
\centering
\caption{Domain size and number of grid points for the R$\rightarrow$S and S$\rightarrow$R LES cases. The grid sizes shown in the parenthesis are for the precursor domain. The number of points in the main domain is $[nx\times ny\times (nz+nz_p)]$}
\label{tab:domain_size}
\begin{tabular}{cccc}
\toprule
Transition type & $(L_x \times L_y \times L_z)/\delta_0$ & ($nx\times ny \times nz$) & $nz_p$\\
\midrule
R$\rightarrow$S & (9.6, 1.6, 1) & (240, 80, 80) & 80\\
S$\rightarrow$R & (43.3, 1.6, 1) & (1024, 80, 80) & 80\\
\bottomrule
\end{tabular}
\end{table}

The upstream friction velocity ($u_{*1}$) and the boundary layer height ($\delta_0$) are used as the velocity and length scales, respectively, for normalization. The origin of the coordinate system is placed at the location of the roughness transition. Each simulation is carried out for $100\,\delta_0/u_{*1}$ and statistical averaging is performed over the last $60\,\delta_0/u_{*1}$. Since the setup is periodic in the spanwise direction, averaging is performed over time as well as along the spanwise direction.

\begin{figure}[h!]
    \centering
    \subfigure[]{\includegraphics[width=0.7\linewidth]{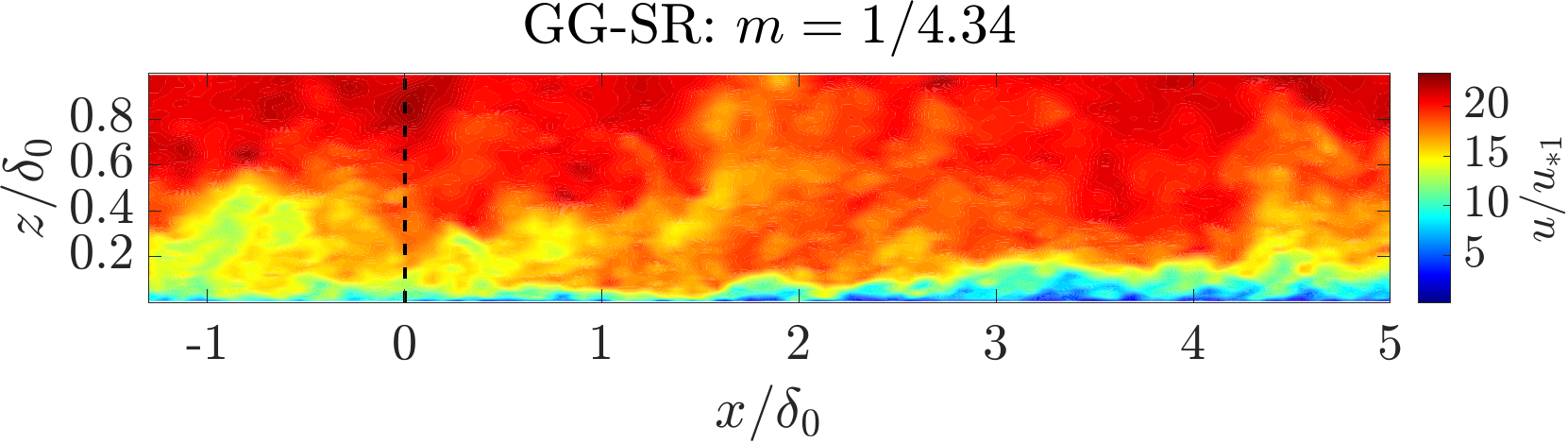}}
    \subfigure[]{\includegraphics[width=0.7\linewidth]{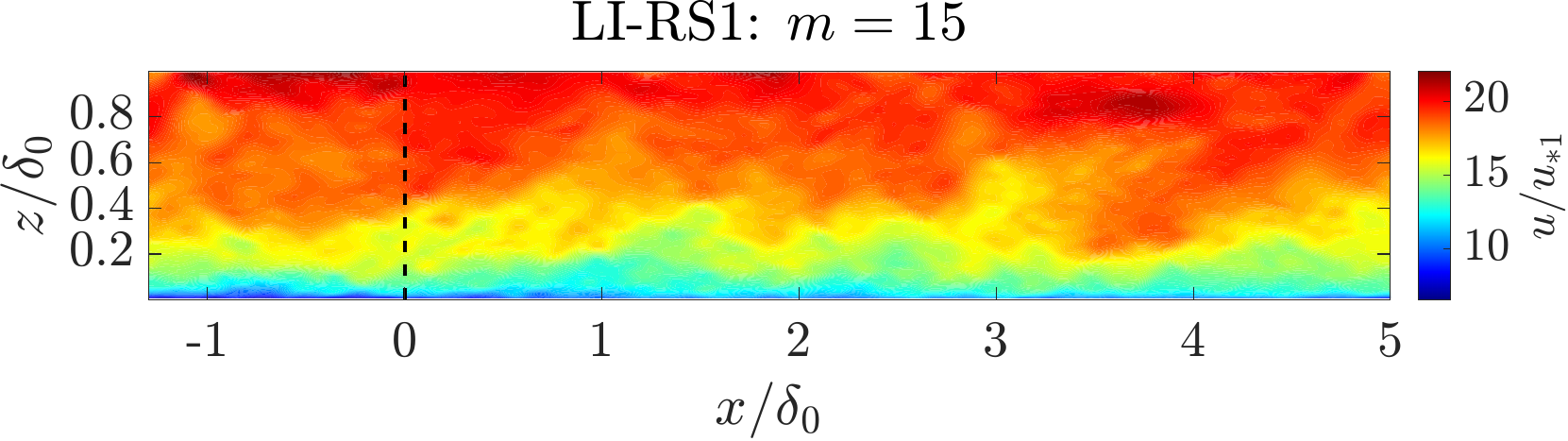}}
    \caption{Instantaneous contours of the mean streamwise velocity ($u$) normalized by the upstream friction velocity ($u_{*1}$) for (a) GG-SR and (b) LI-RS1. The vertical black dashed line represents the roughness transition}
    \label{fig:les_contours}
\end{figure}

\begin{figure}
    \centering
    \includegraphics[width=0.7\linewidth]{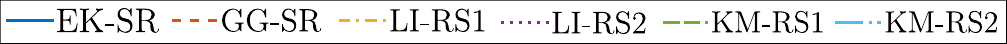}
    \vspace{0.3cm}
    \subfigure[]{\includegraphics[width=0.49\linewidth]{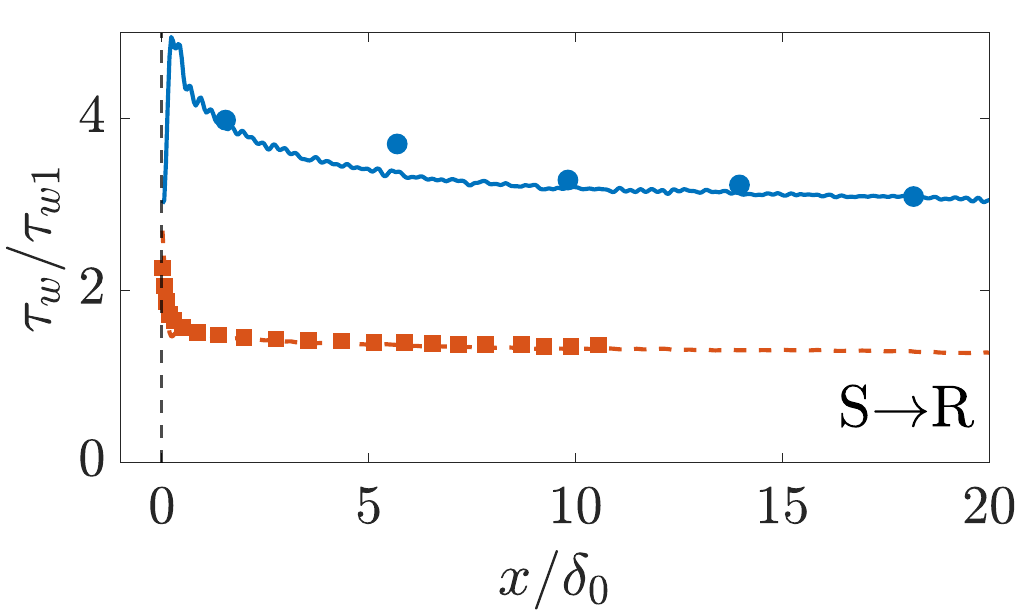}}
    \subfigure[]{\includegraphics[width=0.49\linewidth]{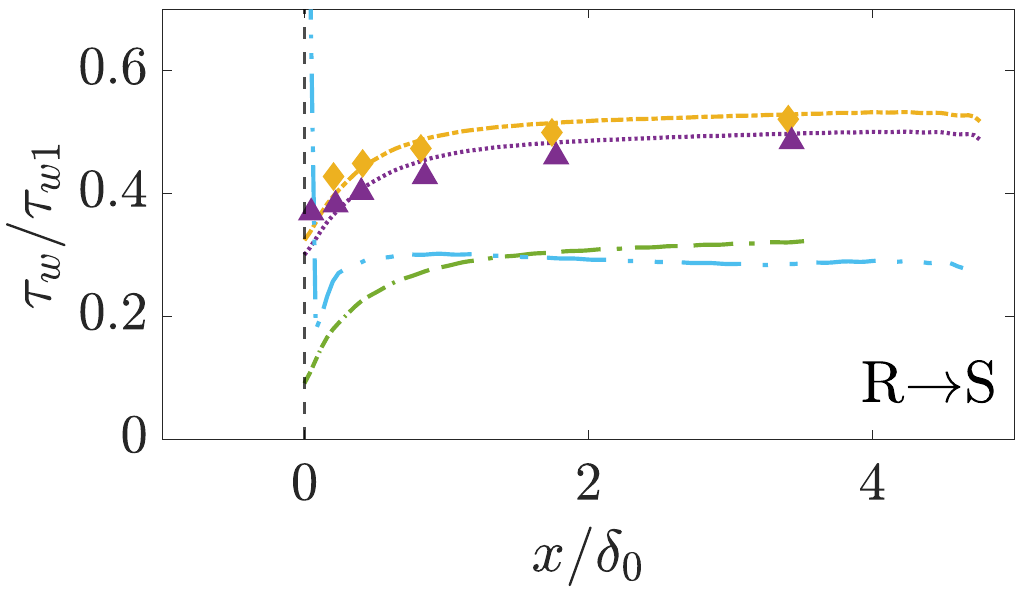}}
    \subfigure[]{\includegraphics[width=0.49\linewidth]{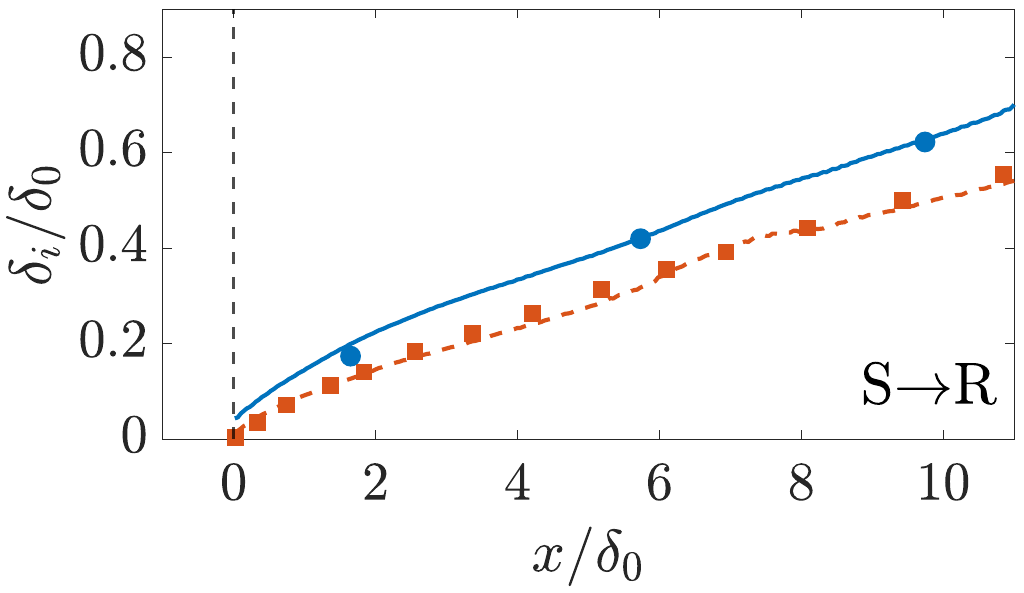}}
    \subfigure[]{\includegraphics[width=0.49\linewidth]{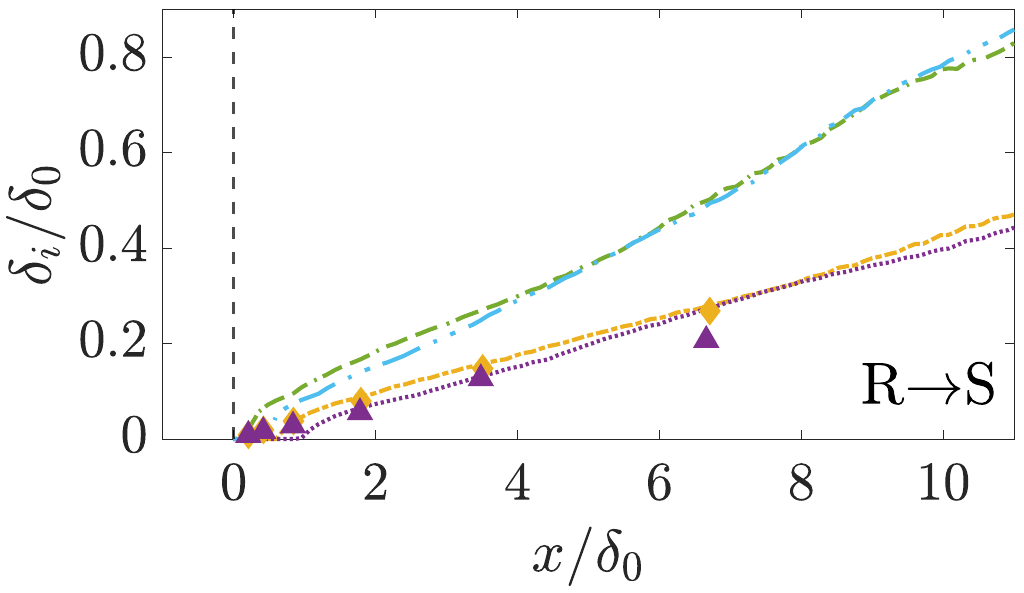}}
    \caption{Streamwise evolution of (a, b) the wall shear stress, $\tau_w=-u_*^2$, downstream of the roughness transition normalized by the upstream wall shear stress, $\tau_{w1}=-u_{*1}^2$, and (c, d) the IBL height, $\delta_i$, normalized by the upstream boundary layer height, $\delta_0$, for cases listed in Table~\ref{tab:cases_train}. The vertical black dashed line represents the roughness transition. Experimental data are shown using markers defined in Table~\ref{tab:cases_train} and LES data are shown using lines defined in the legend}
    \label{fig:les_tauw}
\end{figure}

Instantaneous contours of the streamwise velocity for a representative S$\rightarrow$R and an R$\rightarrow$S transition are shown in Fig.~\ref{fig:les_contours}(a) and \ref{fig:les_contours}(b), respectively, for cases GG-SR and LI-RS1. A clear deceleration (acceleration) close to the ground due to the rougher (smoother) wall is seen downstream of $x=0$ in Fig.~\ref{fig:les_contours}(a) (Fig.~\ref{fig:les_contours}(b)). The streamwise evolution of the wall shear stress, $\tau_w=-u_{*}^2$, for cases listed in Table~\ref{tab:cases_train} is presented in Fig.~\ref{fig:les_tauw}(a) and \ref{fig:les_tauw}(b).  
For an S$\rightarrow$R transition, the rougher downstream surface resists the flow to a greater degree, leading to $\tau_w/\tau_{w1}>1$. Conversely, for an R$\rightarrow$S transition, the smoother downstream surface offers lesser resistance to the flow, leading to $\tau_w/\tau_{w1} < 1$. The IBL height at each streamwise location is defined as the height at which the downstream turbulence intensity profile differs from the upstream profile by less than 10\% \cite{mondal2023r2s}. The streamwise evolution of the IBL is shown in Fig.~\ref{fig:les_tauw}(c) and \ref{fig:les_tauw}(d) for the S$\rightarrow$R and R$\rightarrow$S cases, respectively. Good agreement is observed between the LES and the experiments for $\tau_w$ as well as $\delta_i$. 

We note that a more thorough validation for cases EK-SR, GG-SR, LI-RS2 and KM-RS1 has been reported previously \cite{mondal2023r2s, mondal2025s2r}. LES corresponding to the LI-RS1 case was not reported previously and the agreement with experimental data for this case is shown in this paper for the first time. These well-validated LES results are utilized to develop an analytical modelling framework for the mean velocity, wall shear stress and IBL height, as detailed in the next section. 

%% file: files/blm2020.tex
A schematic of the problem studied is shown in Fig.~\ref{fig:schematic}, with the mean wind along the direction of increasing $x$ and the aerodynamic roughness length undergoing an abrupt transition from $z_{01}$ to $z_{02}$ at $x=0$. The flow upstream of $x=0$ is assumed to be in equilibrium with the surface friction velocity $u_{*1}$ and the resulting mean velocity profile is given by the logarithmic law of the wall, $\bar{u}(z)=(u_{*1}/\kappa)\ln(z/z_{01})$, where $\kappa=0.41$ is the von-K\'arm\'an constant. Downstream of the roughness transition, the analytical model proposed by \citet{ghaisasblm2020} postulates a three-layer structure \cite{abkar2012new} as shown in Fig.~\ref{fig:schematic}. This model derivation is reviewed here briefly. Within the EBL ($z_{02}<z<\delta_e$), the flow is fully adjusted to the downstream surface conditions. The surface friction velocity evolves with the downstream distance and is denoted by $u_{*2,\text{loc}}(x)$. The region between the IBL and the EBL ($\delta_e \leq z < \delta_i$) is called the transition region, where the flow is modified by the changed surface roughness but has not fully adjusted to the downstream conditions.

For brevity of notation, we use a height-dependent `friction velocity' \cite{segalini2023asymptotic,kotturshettar2025mean} $u_{*}(x,z) = \sqrt{-\tau(x,z)}$, where $\tau(x,z)$ is the local vertical shear stress. This friction velocity downstream of the roughness transition is assumed to be vertically invariant in the two equilibrium layers (within the EBL and beyond the IBL) and is assumed to vary linearly with height within the transition region,
\begin{equation}
u_*(x,z)=
\begin{cases}
\displaystyle u_{*2,\mathrm{loc}}(x), & z_{02} \le z < \delta_e(x), \\[6pt]
\displaystyle u_{*2,\mathrm{loc}}(x) + \xi(x)\left(u_{*1}-u_{*2,\mathrm{loc}}(x)\right), & \delta_e(x) \le z < \delta_i(x), \\[6pt]
\displaystyle u_{*1}, & z \ge \delta_i(x).
\end{cases}
\label{eq:ust_piecewise}
\end{equation}

Here, $\xi(x)=(z-\delta_e(x))/(\delta_i(x)-\delta_e(x))$ is a scaled vertical coordinate that becomes 0 at $z=\delta_e$ and 1 at $z=\delta_i$.

The eddy viscosity is also modelled \cite{ghaisasblm2020} as a piecewise function of the wall normal distance,
\begin{equation}
\nu_t(x,z)=
\begin{cases}
u_{*2,\mathrm{loc}}(x)\,\kappa z,
& z_{02} \le z < \delta_e(x), \\

\nu_t^{\mathrm{trans}}(x,\xi),
& \delta_e(x) \le z < \delta_i(x), \\

u_{*1}\kappa z,
& z \ge \delta_i(x) .
\end{cases}
\label{eq:nut_piecewise}
\end{equation}
The eddy viscosity varies linearly with vertical height in the two equilibrium layers. In the transition layer, the eddy viscosity is written as a summation of an equilibrium base profile with an augmentation,
\begin{equation}
    \nu_t^\mathrm{trans} = \nu_t^e +\nu_t^a.
\label{eq:nut}
\end{equation}
The base profile is linear in $z$ and merges with the $\nu_t$ values at the two ends ($\delta_e$ and $\delta_i$) of the transition layer, 
\begin{equation}
    \nu_t^e(x,\xi) = u_{*2,\mathrm{loc}}\kappa\delta_e + (u_{*1}\kappa\delta_i - u_{*2,\mathrm{loc}}\kappa\delta_e)\xi.
\label{eq:nute}
\end{equation}
Downstream of a roughness transition, flow adjustment to the changed surface happens via enhanced vertical mixing, which is modelled here through an augmented eddy viscosity, $\nu_t^a$. A similar approach was reported by \citeauthor{calaf2010large} \cite{calaf2010large}, in which an augmentation was added to the background equilibrium eddy viscosity within the rotor region to account for enhanced mixing. We assume the augmentation to be parabolic in $\xi$,
\begin{equation}
    \nu_t^a(x,\xi) = a\xi^2 + b\xi + c.
\label{eq:nuta1}
\end{equation} 
Three conditions are needed to specify the coefficients $a$, $b$ and $c$ of Eq.~\eqref{eq:nuta1}. Two of these conditions are obtained by imposing continuity at $z=\delta_e$ and $\delta_i$, which requires $\nu_t^a(x,0) = \nu_t^a(x,1) = 0$. The third constraint is chosen to be $\nu_t^a(x,0.5) = \beta \nu_\mathrm{mid}$, where $\nu_\mathrm{mid}=(u_{*2,\mathrm{loc}}\kappa\delta_e+u_{*1}\kappa\delta_i)/2$ is the base profile value at the midpoint of the transition layer, and $\beta$ is a free parameter that controls the strength of the augmentation. Using these constraints and evaluating the coefficients $a$, $b$ and $c$, the resulting form for the eddy viscosity augmentation is 
\begin{equation}
    \nu_t^a(x,\xi) = 2\beta\kappa\xi(1-\xi)(u_{*1}\delta_i+u_{*2,\mathrm{loc}}\delta_e).
\label{eq:nuta}
\end{equation}
Putting Eqs.~\eqref{eq:nute} and  \eqref{eq:nuta} in Eq.~\eqref{eq:nut} and rearranging leads to 
\begin{equation}
\label{eq:nut_trans}
    \nu_t^\mathrm{trans}(x,\xi) = -2\beta\kappa(u_{*2,\mathrm{loc}}\delta_e+u_{*1}\delta_i)\xi^2 + [u_{*1}\delta_i(1+2\beta)-u_{*2,\mathrm{loc}}\delta_e(1-2\beta)]\kappa\xi + u_{*2,\mathrm{loc}}\kappa\delta_e.
\end{equation}

Expressions for the mean velocity gradients in the three layers can be derived using Eqs. \eqref{eq:ust_piecewise}, \eqref{eq:nut_piecewise} and \eqref{eq:nut_trans}. Within the EBL, i.e., $z_{02}\leq z < \delta_e$, the velocity gradient is given by
\begin{equation}
    \frac{\partial \bar{u}}{\partial z} = \frac{-\tau}{\nu_t} = \frac{u_{*2,\mathrm{loc}}}{\kappa z},
\end{equation}
integrating which from $z=z_{02}$ to $z$ gives
\begin{equation}
    \bar{u}(z) = \frac{u_{*2,\mathrm{loc}}}{\kappa}\ln\left(\frac{z}{z_{02}}\right).
    \label{eq:u_eqb_lower}
\end{equation}
Similarly, for the region above the IBL ($z\geq \delta_i$), integrating from $\delta_i$ to $z$ gives
\begin{equation}
    \bar{u}(z) = \frac{u_{*1}}{\kappa}\ln\left(\frac{z}{z_{01}}\right), \label{eq:u_eqb_upper}
\end{equation}
which is the logarithmic law of the wall for the upstream conditions, consistent with the assumption that the flow here is unaffected by the changed surface conditions and still in equilibrium with the upstream conditions. To evaluate the velocity gradient in the transition layer, two cases must be considered depending on the value of $\beta$. For $\beta \neq 0$, the velocity gradient in the transition layer is
\begin{equation}
    \frac{\partial \bar{u}}{\partial z} = \frac{-\tau}{\nu_t} = \frac{(u_{*1}-u_{*2,\mathrm{loc}})^2}{-2\beta\kappa(u_{*2,\mathrm{loc}}\delta_e+u_{*1}\delta_i)} \left[ \frac{\xi^2+p\xi+q}{\xi^2+r\xi+s} \right]
\end{equation}
where $p$, $q$, $r$ are $s$ are algebraic functions of $u_{*1}$, $u_{*2,loc}$, $\delta_i$, $\delta_e$ and $\beta$ (detailed derivation is given in Appendix~\ref{app:utrans}). Integrating from the top, $\xi'=1$, to an arbitrary $\xi$ and imposing continuity with the equilibrium profile at $\xi=1$ gives,
\begin{align}
\bar{u}(\xi) =\;&
\frac{u_{*1}}{\kappa}
\ln\left(\frac{\delta_i}{z_{01}}\right)
\nonumber + \frac{\left(u_{*1}-u_{*2,\mathrm{loc}}\right)^2
(\delta_i-\delta_e)}
{-2\beta\kappa
\left(
u_{*2,\mathrm{loc}}\delta_e
+
u_{*1}\delta_i
\right)}
\nonumber 
\Bigg[
\xi-1
+
\frac{p-r}{2}
\ln\left(
\frac{\xi^2+r\xi+s}
{1+r+s}
\right)
\nonumber \\[6pt]
&\qquad
+
\frac{
q+r^2/2-pr/2-s
}
{2\mathcal{R}}
\ln\left|
\frac{
(\xi+r/2-\mathcal{R})(1+r/2+\mathcal{R})
}{
(\xi+r/2+\mathcal{R})(1+r/2-\mathcal{R})
}
\right|
\Bigg].
\label{eq:u_trans}
\end{align}
Here, $\mathcal{R}=\sqrt{r^2/4-s}$. The velocity profile in the transition region for $\beta=0$ is shown in Appendix~\ref{app:utrans}. Equating Eq.~\eqref{eq:u_trans} at $\xi=0$ to the equilibrium velocity at $z=\delta_e$ enforces continuity of mean velocity across the EBL and yields a non-linear equation with two unknowns ($u_{*2,loc}$ and $\delta_i$),
\begin{align}
\frac{u_{*2,loc}}{\kappa}
\ln\left(\frac{\delta_e}{z_{02}}\right)
 =\;&
\frac{u_{*1}}{\kappa}
\ln\left(\frac{\delta_i}{z_{01}}\right)
\nonumber + \frac{\left(u_{*1}-u_{*2,\mathrm{loc}}\right)^2
(\delta_i-\delta_e)}
{-2\beta\kappa
\left(
u_{*2,\mathrm{loc}}\delta_e
+
u_{*1}\delta_i
\right)}
\nonumber 
\Bigg[
-1
+
\frac{p-r}{2}
\ln\left(
\frac{s}
{1+r+s}
\right)
\nonumber \\[6pt]
&\qquad
+
\frac{
q+r^2/2-pr/2-s
}
{2\mathcal{R}}
\ln\left|
\frac{
(r/2-\mathcal{R})(1+r/2+\mathcal{R})
}{
(r/2+\mathcal{R})(1+r/2-\mathcal{R})
}
\right|
\Bigg].
\label{eq:u_model}
\end{align}

Model closure requires a-priori specification of the IBL height, along with suitable choices of the parameters $\alpha$ and $\beta$. In the work of \citet{ghaisasblm2020}, the empirical expression for $\delta_i$ proposed by \citet{elliott1958growth} was used. This empirical expression, however, has been shown to be inaccurate in several R$\rightarrow$S and S$\rightarrow$R transition studies \cite{mondal2023r2s, mondal2025s2r}. To further improve this model and provide a self-consistent physics-based framework, we develop an analytical model for $\delta_i$ in the next section. The equation derived in the next section can be coupled with Eq.~\eqref{eq:u_model} to determine $u_{*2,loc}$ and $\delta_i$ simultaneously.

%% file: files/ibl_model.tex
A surface roughness transition in the ABL flow induces streamline displacement, acceleration or deceleration in the mean streamwise velocity, and an imbalance between turbulence production and dissipation \cite{ding2025ibl}. \citet{miyake1965} hypothesized that these imbalances diffuse upwards with the strength of diffusion measured by the standard deviation of the vertical velocity fluctuations, $\sigma_w=\sqrt{\overline{w^\prime w^\prime}}$, and hence related the rate of growth of the IBL to be proportional to $\sigma_w$. The study by \citet{savelyev2005internal} further incorporated the effects of streamline displacement by adding a term proportional to the mean vertical velocity, $\bar{w}$, via the equation 
\begin{equation}
\label{eq:diffusion_eq}
    \frac{\text{d}\delta_i(x)}{\text{d}t} = A_1\sigma_w(x,\delta_i) +A_2 \bar{w}(x,\delta_i), 
\end{equation}
where $A_1$ and $A_2$ are proportionality factors. To simplify the left-hand side of this equation, following \citet{savelyev2005internal}, $\text{d}\delta_i/\text{d}t$ is expressed as $\bar{u}(\text{d}\delta_i/\text{d}x)$ using the chain rule. At the IBL height, $\bar{u}$ is calculated using the logarithmic law of the wall adjusted to the upstream surface to maintain continuity of velocity. Thus, the equation of evolution of the IBL becomes
\begin{equation}
\label{eq:chain_rule}
    \left[\frac{u_{*1}}{\kappa}\ln{\left(\frac{\delta_i}{z_{01}}\right)}\right]\frac{\text{d}\delta_i}{\text{d}x} = A_1\sigma_w(x,\delta_i) +A_2 \bar{w}(x,\delta_i).
\end{equation}
The simplification of the right-hand side of this equation from hereon differs from the assumptions and choices made by \citet{savelyev2005internal}. We highlight these differences and provide justifications for our choices below. There are two major points of departure as detailed below.
\subsubsection{Turbulent diffusion of momentum}
As noted by \citet{savelyev2005internal}, the first term of the right-hand side of Eq.~\eqref{eq:diffusion_eq} represents the diffusion of excess or deficit of momentum, similar to the diffusion from a source of a passive scalar due to turbulence. 
Following this diffusion analogy of \citet{miyake1965}, \citet{savelyev2005internal} assumed that the vertical diffusion intensity, $\sigma_w$, is proportional to the upstream friction velocity, $u_{*1}$. However, this assumption leads to identical IBL growth rates for cases where the upstream roughness, $z_{01}$, is the same but the downstream, $z_{02}$, values differ, which is physically incorrect. As the strength of the upward diffusion is induced by the new surface conditions, $\sigma_w$ should predominantly depend on the downstream friction velocity, $u_{*2,loc}$. This was also discussed by \citet{panofskyDutton1984}, and following their study, we adopt $\sigma_w=C \:u_{*2,loc}$ with $C=1.25$.  

\begin{figure}
    \centering
    \subfigure[]{\includegraphics[width=0.325\linewidth]{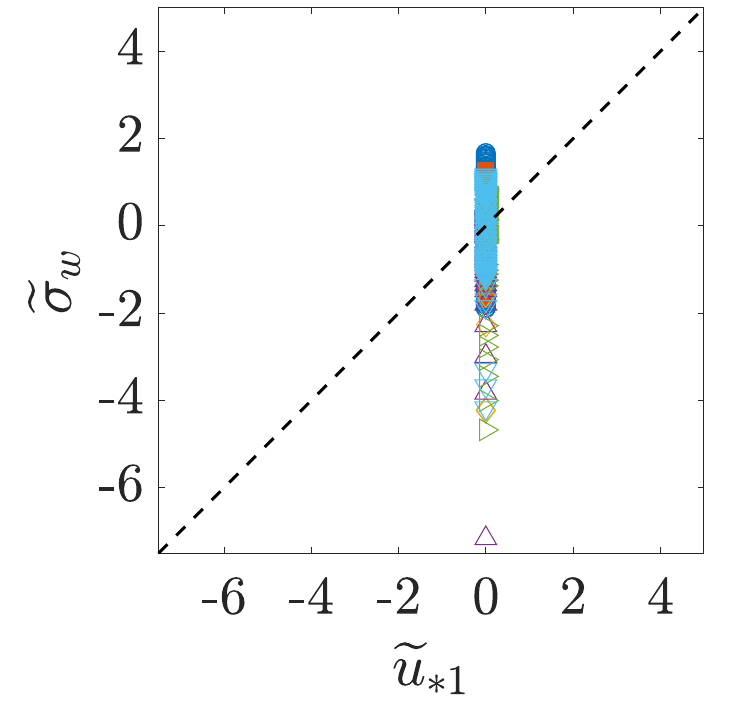}}
    \subfigure[]{\includegraphics[width=0.325\linewidth]{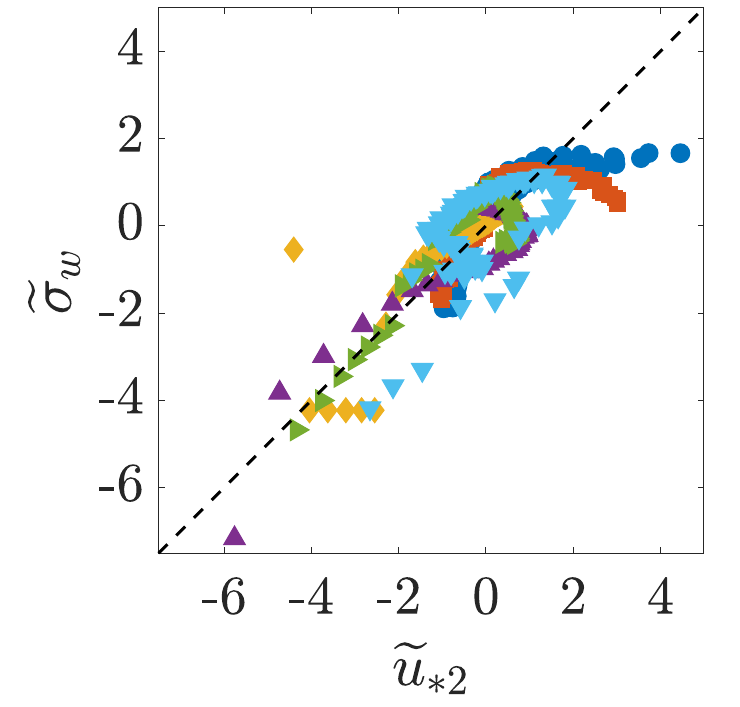}}
    \subfigure[]{\includegraphics[width=0.325\linewidth]{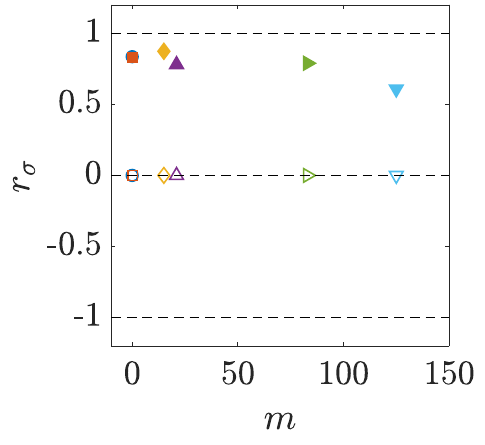}}
    \caption{Scatter plot of standardized velocity variance ($\widetilde{\sigma}_w$) with the standardized (a) upstream, $\widetilde{u}_{*1}$,  and (b) downstream, $\widetilde{u}_{*2,loc}$, friction velocities. The diagonal black dashed line has slope 1 and passes through the origin. (c) Correlation coefficient $r_{\sigma}$ for the six LES cases shown in Table~\ref{tab:cases_train} plotted as a function of roughness ratio, $m$. Symbol shapes and corresponding LES cases are in accordance with Table~\ref{tab:cases_train}. In panel (c), the hollow symbols show $r(\sigma_w,u_{*1})$ corresponding to the ST model while the filled symbols show $r(\sigma_w,u_{*2,loc})$ corresponding to the new model assumption}
    \label{fig:sigma_corr}
\end{figure}

To evaluate the assumptions of the ST model and our proposed model, the standardized variables for $\sigma_w$ and the two friction velocities are compared in Fig.~\ref{fig:sigma_corr}. The standardized variable for any variable $\chi$ is given by
\begin{equation}
    \widetilde{\chi} = \frac{\chi-\overline{\chi}}{\sqrt{\mathrm{Var}(\chi)}},
\end{equation}
where ( $\overline{\vphantom{int}...}$ ) denotes the mean of the variable and the operator `Var' denotes its variance. 
The LES data from all the cases listed in Table~\ref{tab:cases_train} is used to compute $\widetilde{\sigma}_w$,  $\widetilde{u}_{*1}$ and $\widetilde{u}_{*2,loc}$. For each case, the mean and variance are computed using these values at different streamwise locations, $x$, behind the roughness transition. Since $u_{*1}$ is a constant (i.e. independent of $x$) in each case, its standardized variable $\widetilde{u}_{*1}$ is set to zero. 

Figure~\ref{fig:sigma_corr}(a) shows that the $\widetilde{\sigma}_w$ values do not correlate with the $\widetilde{u}_{*1}$ values across all cases. On the other hand, the data points cluster around the diagonal (black dashed) line with slope 1 in Fig.~\ref{fig:sigma_corr}(b), indicating a fairly strong linear relationship between $\sigma_w$ and $u_{*2,loc}$. Figure~\ref{fig:sigma_corr}(c) further quantifies these dependencies by plotting the Pearson correlation coefficient between $\sigma_w$ and the two friction velocities. The Pearson correlation coefficient between two variables $\chi$ and $\phi$ is given by
\begin{equation}
    \label{eq:corr_coeff}
    r(\chi,\phi) =  \frac{\sum\limits_{i=1}^N(\chi_i-\overline{\chi})(\phi_i-\overline{\phi})}{\sqrt{\sum\limits_{i=1}^N(\chi_i-\overline{\chi})^2 \sum\limits_{i=1}^N(\phi_i-\overline{\phi})^2}}=\frac{1}{N}\sum\limits_{i=1}^N\widetilde{\chi}_i\widetilde{\phi}_i,
\end{equation}
where $N$ is the number of data points (here, the number of downstream locations $x$) in a given case. The correlation coefficient is, by definition, bounded within $[-1, \; 1]$, and $r=0$ denotes no correlation while $r=1$ and $-1$ indicate perfect positive and negative correlations, respectively. 
The hollow symbols in Fig.~\ref{fig:sigma_corr}(c) represent $r(\sigma_w, u_{*1})$, while the filled symbols represent $r(\sigma_w, u_{*2,\mathrm{loc}})$. Figure~\ref{fig:sigma_corr}(c) clearly demonstrates that there is no correlation between $\sigma_w$ and $u_{*1}$, while a strong positive correlation, between 0.6 to 0.9, exists between $\sigma_w$ and $u_{*2,\mathrm{loc}}$. Thus, the assumption in our proposed model is physically justified and $\sigma_w$ is strongly dependent on the friction velocity downstream of the roughness transition.
\subsubsection{Streamline displacement}
The second term on the right-hand side of Eq.~\eqref{eq:diffusion_eq} represents the effects of streamline displacement caused by an accelerating or decelerating streamwise flow. To model the streamline displacement, the mean vertical velocity at the IBL height can be approximated as $\bar{w}\approx - \Delta \bar{u} \:\delta_i/x$, using the continuity equation in a weak sense \citep{savelyev2005internal}, where $\Delta \bar{u}$ represents the characteristic difference between the local mean streamwise velocity at $z=\delta_i$ and the corresponding value upstream of the roughness change where the vertical velocity is zero. In this study, we approximate this characteristic velocity difference as
\begin{equation}
    \Delta \bar{u} = \frac{u_{*2,loc}}{\kappa}\ln\left(\frac{\delta_i}{z_{02}}\right)-\frac{u_{*1}}{\kappa}\ln\left(\frac{\delta_i}{z_{01}}\right). \label{eq:delu}
\end{equation}
Rearranging the expression for $\Delta u$ and putting it in the expression for $\bar{w}(x)$ yields
\begin{equation}
\label{eq:vert_vel}
    \bar{w}=-\Delta \bar{u}\frac{\delta_i}{x} = \left[\frac{(u_{*1}-u_{*2,loc})}{\kappa} \ln\left(\frac{\delta_i}{z_{01}}\right)+\frac{u_{*2,loc}}{\kappa}\ln\left(\frac{1}{m}\right)\right]\frac{\delta_i}{x}.
\end{equation}
In contrast, \citet{savelyev2005internal} argued that, at far-downstream locations, the wall shear stress becomes comparable to the upstream wall shear stress, and hence replaced $u_{*2,loc}$ with $u_{*1}$ in the above expressions to yield
\begin{equation}
\label{eq:vert_vel_ST}
    \bar{w}=-\Delta \bar{u}\frac{\delta_i}{x} = \left[\frac{u_{*1}}{\kappa}\ln\left(\frac{1}{m}\right)\right]\frac{\delta_i}{x}.
\end{equation}

Several previous studies \citep{efros2011s2r,li2021r2s,mondal2023r2s,mondal2025s2r} on ABL flow over heterogeneous surfaces have shown that the wall shear stress varies across a roughness transition, and $\tau_w/\tau_{w1}$ asymptotes to a higher value over an S$\rightarrow$R transition and a lower value over an R$\rightarrow$S transition, as also seen in Fig.~\ref{fig:les_tauw}. Thus, the assumption $u_{*2,loc}\approx u_{*1}$ is incorrect. Furthermore, using this assumption in Eq.~\eqref{eq:delu} leads to the characteristic difference, $\Delta \bar{u}$, to be independent of $x$ and would lead to either an overestimation or an underestimation of the IBL height. 

\begin{figure}
    \centering
    \subfigure[]{\includegraphics[width=0.325\linewidth]{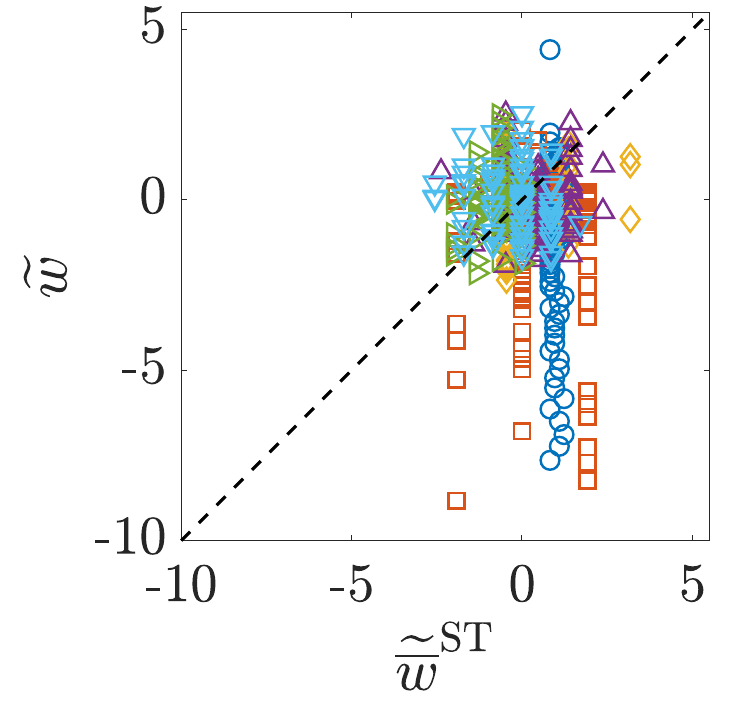}}
    \subfigure[]{\includegraphics[width=0.325\linewidth]{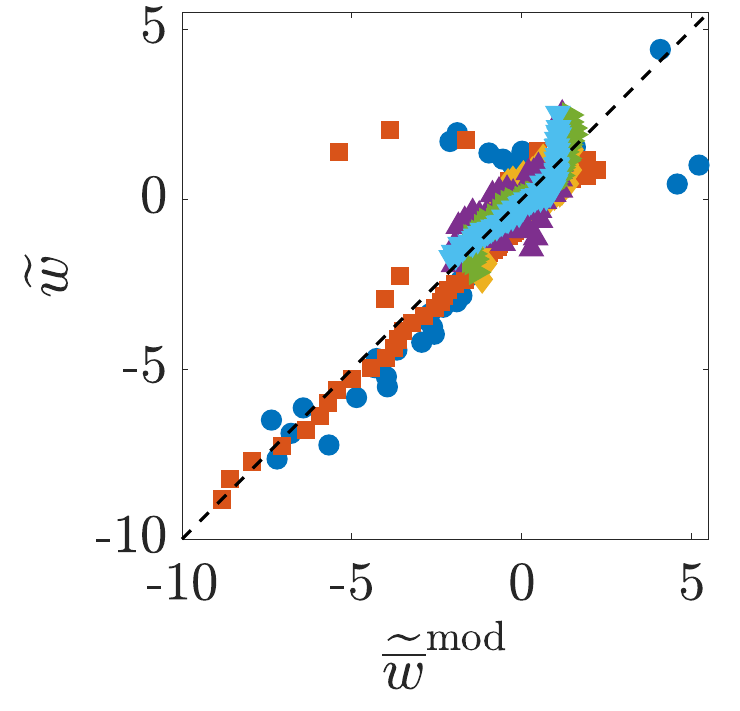}}
    \subfigure[]{\includegraphics[width=0.325\linewidth]{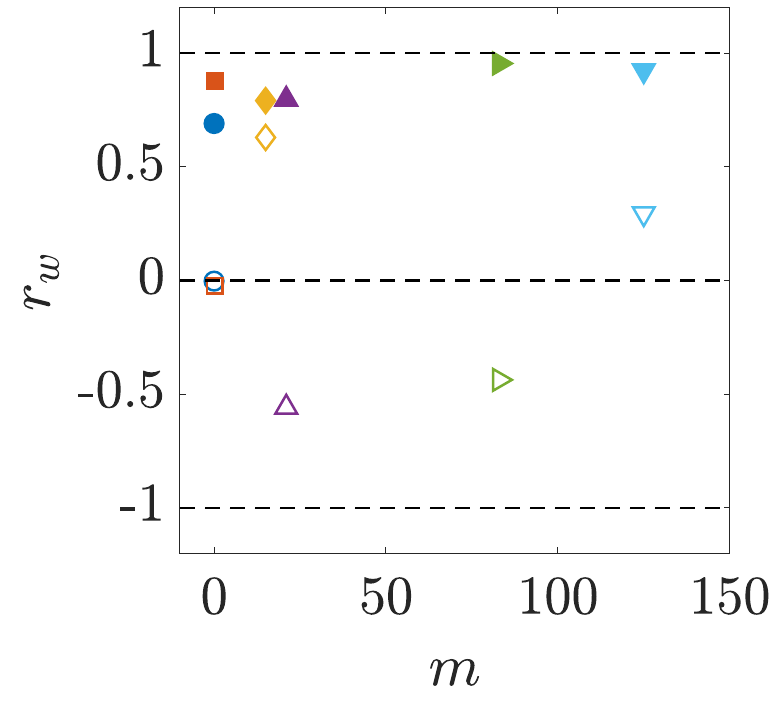}}
    \caption{Scatter plot of the standardized vertical velocity obtained from the LES data, $\widetilde{\bar{w}}^{\mathrm{LES}}$, and the standardized vertical velocity predicted by (a) the ST model, $\widetilde{\bar{w}}^{\mathrm{ST}}$, Eq.~\eqref{eq:vert_vel_ST}, and (b) the current proposed model, $\widetilde{\bar{w}}^{\mathrm{mod}}$, Eq.~\eqref{eq:vert_vel}. The diagonal black dashed line has slope 1 and passes through the origin. (c) Correlation coefficient $r_{w}$ for the six LES cases shown in Table~\ref{tab:cases_train} plotted as a function of roughness ratio, $m$. In panel (c), the hollow symbols show $r(\bar{w}^{\mathrm{LES}}, \bar{w}^{\mathrm{ST}})$ corresponding to the ST model while the filled symbols show $r(\bar{w}^{\mathrm{LES}}, \bar{w}^{\mathrm{mod}})$ corresponding to our current proposed model}
    \label{fig:w_corr}
\end{figure}

The two choices for modelling $\bar{w}$ are evaluated in   Fig.~\ref{fig:w_corr}. The actual vertical velocity, $\bar{w}^{LES}$, and the vertical velocities predicted by the ST model (Eq.~\eqref{eq:vert_vel_ST}) and our new proposed model (Eq.~\eqref{eq:vert_vel}) are extracted using data at multiple downstream locations from the six LES cases listed in Table~\ref{tab:cases_train}.  
The standardized variables corresponding to these are plotted in Fig.~\ref{fig:w_corr}(a) and \ref{fig:w_corr}(b). Most of the data points in Fig.~\ref{fig:w_corr}(a) cluster around (0,0) indicating a lack of correlation between $\bar{w}^{LES}$ and the ST model. This can also be seen quantitatively in Fig.~\ref{fig:w_corr}(c), where the hollow symbols display a range of correlation coefficient values ranging from -0.5 to 0.5. 
On the other hand, the standardized variable data points corresponding to the actual vertical velocity and our new model predictions largely fall on the diagonal line with a slope of 1 in Fig.~\ref{fig:w_corr}(b). Consequently, the correlations for our proposed model (filled symbols in Fig.~\ref{fig:w_corr}(c)) are close to unity across most cases. These results suggest that the assumptions of the present model are physically consistent with the LES data and that it improves over the previous ST model considerably.

\begin{figure}
    \centering
    \includegraphics[width=0.5\linewidth]{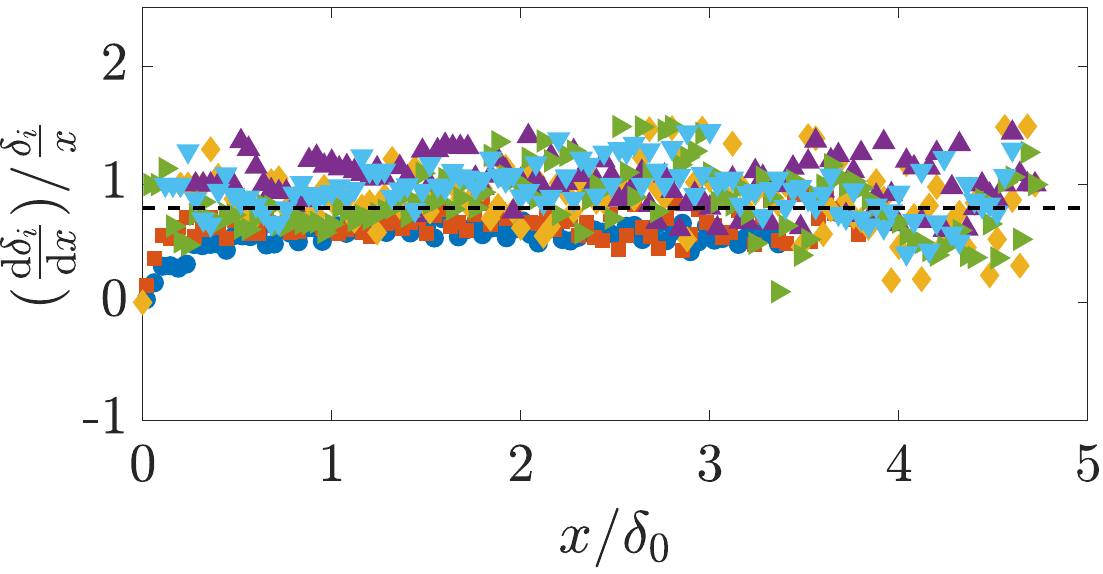}
    \caption{Streamwise evolution of the ratio of the IBL growth rate ($\mathrm{d}\delta_i/\mathrm{d}x$) and the ratio of IBL height normalized by the downstream distance ($\delta_i/x$). The black dashed horizontal line indicates a ratio of 0.8. The symbols indicate the different cases as shown in Table~\ref{tab:cases_train}}
    \label{fig:delifx}
\end{figure}

Substituting Eq.~\eqref{eq:vert_vel} and the revised expression for vertical velocity variance, $\sigma_w=Cu_{*2,loc}$, in Eq.~\eqref{eq:chain_rule} gives
\begin{equation}
\label{eq:delta_eq_stg1}
\left[\frac{u_{*1}}{\kappa}\ln{\left(\frac{\delta_i}{z_{01}}\right)}\right]\frac{\text{d}\delta_i}{\text{d}x} = A_1 C u_{*2,loc} + A_2 \left[\frac{(u_{*1}-u_{*2,loc})}{\kappa} \ln\left(\frac{\delta_i}{z_{01}}\right)+\frac{u_{*2,loc}}{\kappa}\ln\left(\frac{1}{m}\right)\right]\frac{\delta_i}{x}.
\end{equation}
To further simplify Eq.~\eqref{eq:delta_eq_stg1}, we use the data from the six LES cases reported in Table~\ref{tab:cases_train} and examine the relationship between $\text{d}\delta_i/\text{d}x$ and $\delta_i/x$. Figure~\ref{fig:delifx} shows that the ratio of $\text{d}\delta_i/\text{d}x$ and $\delta_i/x$ generally does not increase or decrease with the downstream distance. As a simple model, it is fair to assume $\text{d}\delta_i/\text{d}x \approx n\:(\delta_i/x)$, and the LES data (Fig.~\ref{fig:delifx}) show that the proportionality constant, $n$, lies in the range $0.6\leq n \leq 0.9$. We adopt $n=0.8$, indicated by the black dashed line in Fig.~\ref{fig:delifx}, obtained by averaging over all the LES data. It is worth noting that a power law of the form $\delta_i\sim x^{0.8}$, which is consistent with $n=0.8$ that we adopt here, has been assumed in several previous empirical models of IBL height reported in the literature \cite{elliott1958growth,wood1982internal,pendergrass1984dispersion,jegede1999study}.
Substituting for $\text{d}\delta_i/\text{d}x$ and simplifying Eq.~\eqref{eq:delta_eq_stg1}, we arrive at
\begin{equation}
\label{eq:delta_model}
\delta_i \left[ \{u_{*1}(n-A_2)+A_2 u_{*2,loc}\}\ln\left(\frac{\delta_i}{z_{01}}\right) - A_2 u_{*2,loc} \ln\left(\frac{1}{m}\right) \right] = A_1 C \kappa u_{*2,loc} x.
\end{equation}
This is a non-linear equation with two unknowns, i.e., $\delta_i$ and $u_{*2,loc}$. 
Together with Eq.~\eqref{eq:u_model}, it forms a system of two non-linear equations that can be solved together to yield $u_{*2,loc}(x)$ and $\delta_i(x)$ at every location $x$ downstream of the roughness transition. These coupled non-linear equations are solved numerically, e.g., using the \texttt{fsolve} function in MATLAB \citep{mathews2004numerical}. The friction velocity and IBL height values are then used along with Eqs.~\eqref{eq:u_eqb_lower}, \eqref{eq:u_eqb_upper} and \eqref{eq:u_trans} to yield the entire profile of the streamwise velocity behind a roughness transition.

Eqs.~\eqref{eq:u_model} and \eqref{eq:delta_model} have four parameters, namely $\alpha$, $\beta$, $A_1$ and $A_2$, at this point. The parameter $\alpha$ is the ratio of the equilibrium boundary-layer height to the internal boundary-layer height, $\alpha=\delta_e/\delta_i$, and was reported to be equal to 0.027 in prior studies \cite{abkar2012new, mondal2023r2s, mondal2025s2r}. The parameter $A_2$ controls the contribution of streamline displacement to the overall IBL growth rate in Eq.~\eqref{eq:diffusion_eq}. \citet{savelyev2005internal} studied several experimental datasets and showed $A_2=C\kappa=0.5$. We retain the same values for these two coefficients for simplicity in our study and find that it is not necessary to further tune these parameters from one case to another. The modeling framework, thus, is dependent on only two parameters, $\beta$ and $A_1$. Comments on specifying these along with an evaluation of the entire modelling framework is discussed in the next section.

%% file: files/results.tex
\subsection{Specification of model coefficients}
\label{sec:coeffs}
The model predictions for wall shear stress, $\tau_w=-u_{*2,loc}^2$, mean velocity, $\bar{u}$, and IBL height, $\delta_i$, are compared with available experimental and LES datasets listed in Table~\ref{tab:cases_train} and \ref{tab:cases_test} in this section. The tunable parameters ($\beta$, $A_1$) need to be specified for model closure. 
To ascertain optimal values of $\beta$ and $ A_1$, predictions are first made over a range of values of each parameter. For each predicted quantity, the $L_2$ norm of the error between the six LES 
datasets (Table~\ref{tab:cases_train}) and the model predictions is calculated. For example, the $L_2$ norm for $\delta_i$ predictions is calculated as  
\begin{equation}
    \epsilon_{\delta} = \sqrt{\frac{1}{N}\sum_{i=1}^N\left(\frac{\delta_i^{\text{LES}}-\delta_i^{\text{model}}}{\delta_{\mathrm{ref}}}\right)^2} \times 100.
\end{equation}
Here, the subscript `$\mathrm{ref}$' denotes the upstream equilibrium value of the corresponding quantity and $N$ is the number of measured data points. To account for errors in all three quantities of interest simultaneously, we introduce a measure for the composite error ($\epsilon_c$),
\begin{equation}
    \epsilon_{c} = \frac{\epsilon_{\tau}}{\text{max}(\epsilon_{\tau})} + \frac{\epsilon_{u}}{\text{max}(\epsilon_{u})} + \frac{\epsilon_{\delta}}{\text{max}(\epsilon_{\delta})}, \label{eq:err}
\end{equation}
where $\epsilon_{\tau}$, $\epsilon_u$ and $\epsilon_{\delta}$ are the $L_2$ norms for $u_{*2,loc}$, $\bar{u}$ and $\delta_i$, respectively. Table~\ref{tab:cases_train_optfit} lists the optimal values (i.e., the combination that yields the minimum $\epsilon_c$) of the model parameters for each case in Set A. Subsequently, a curve fitting is employed on these optimal values so as to generalize the tunable parameters to other cases beyond the six cases listed in Tables~\ref{tab:cases_train} and \ref{tab:cases_train_optfit}. 

\begin{figure}
    \centering
    \subfigure[]{\includegraphics[width=0.36\linewidth]{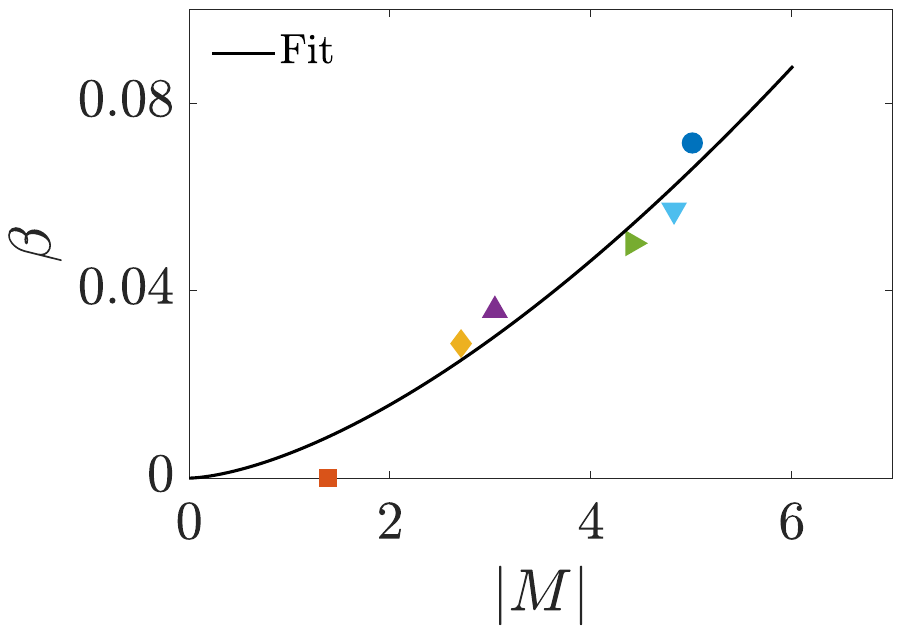}}
    \subfigure[]{\includegraphics[width=0.35\linewidth]{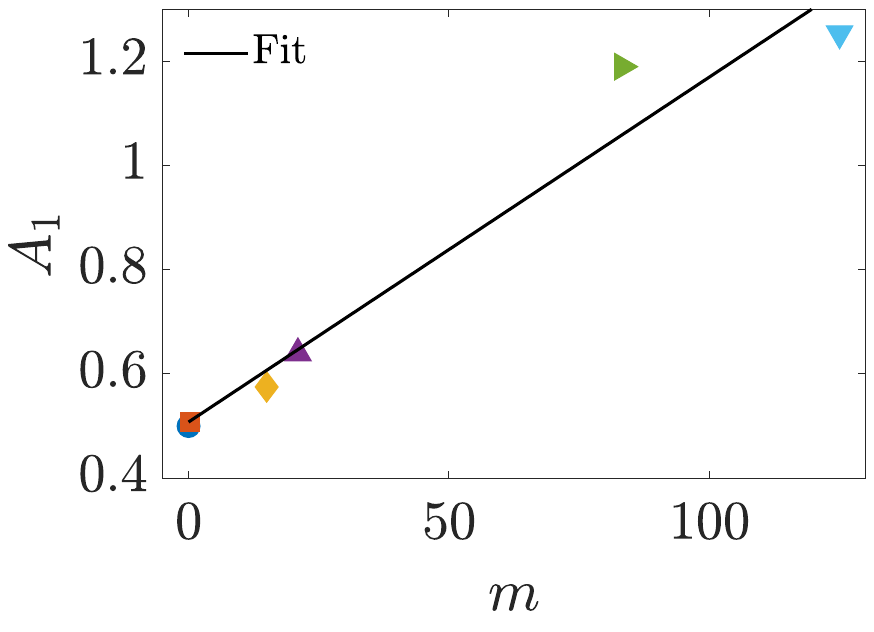}}
    \caption{Optimal coefficient values, (a) $\beta_{\mathrm{opt}}$ and (b) $A_{1,\mathrm{opt}}$, for the six cases listed in Tables~\ref{tab:cases_train} and ~\ref{tab:cases_train_optfit} along with their curve fits, Eqs.~\eqref{eq:betafit} and \eqref{eq:A1fit}. The $R^2$ values for the fits are (a) 0.94 and (b) 0.96}
    \label{fig:A1_beta_fit}
\end{figure}

The parameter $\beta$ controls the non-linearity in the augmentation term of the eddy viscosity (Eq.~\eqref{eq:nut_trans}) in the transition region, $\delta_e \leq z < \delta_i$. The optimal values indicate that $\beta$ depends on the magnitude of $M$, where $M=\ln(1/m)$. The value of $M$ indicates the strength of the roughness jump, and the sign indicates the type of roughness transition (negative for R$\rightarrow$S and positive for S$\rightarrow$R). Accordingly, we model $\beta$ as a function of $|M|$ using a power law, 
\begin{equation}
\label{eq:betafit}
    \beta=0.0053 \:|M|^{1.57}.
\end{equation}
Figure~\ref{fig:A1_beta_fit}(a) shows that this expression provides a reasonable fit to the $\beta_{\mathrm{opt}}$ values of the different cases of Set A. This functional form also ensures that for a homogeneous case ($M=0$) $\beta$ goes to zero, indicating no augmentation to the eddy viscosity on a homogeneously rough surface.  

The IBL growth rate depends on the strength of the discontinuity in surface roughness, as reflected by the roughness ratio. Consequently, $A_1$, the primary parameter governing the IBL growth rate, should be modelled as a function of $m$. The optimal $A_1$ values shown in Table~\ref{tab:cases_train} can be reasonably modelled with a linear fit
\begin{equation}
\label{eq:A1fit}
    A_1 = 0.51+0.0066\:m,
\end{equation} 
as seen in Fig.~\ref{fig:A1_beta_fit}(b). 

The coefficients of determination, $R^2$, that quantify the goodness of fits for the fitted models of $\beta_{\mathrm{opt}}$ and $A_{1,\mathrm{opt}}$ are 0.94 and 0.96, respectively. Thus, while the fits are not perfect, as seen for example at $|M|\approx1.6$ in Fig.~\ref{fig:A1_beta_fit}(a) and around $m=100$ in Fig.~\ref{fig:A1_beta_fit}(b), they provide reasonably accurate representations of the model coefficients and, as a result, of the underlying data. For all subsequent predictions, including those of set A cases, $\beta$ and $A_1$ obtained from Eqs.~\eqref{eq:betafit} and \eqref{eq:A1fit} are used.
\begin{table}[htbp]
\centering
\caption{Optimal coefficient values, $\beta_\text{opt}$ and  $A_{1,\text{opt}}$, that yield the minimum composite error, Eq.~\eqref{eq:err}, for the six LES cases of Set A. Roughness ratio is $m=z_{01}/z_{02}$}
\label{tab:cases_train_optfit}
\makebox[\textwidth][c]{%
\begin{tabular}{ccccc}
\toprule
No. & Label & $m$   & $\beta_\text{opt}$ &  $A_{1,\text{opt}}$\\
\midrule
1 & EK-SR & 1/150 & 0.0735 & 0.49    \\
2 & GG-SR & 1/4.34 & 0.0001 & 0.50 \\
3 & LI-RS1 & 15 & 0.0348 & 0.59 \\
4 & LI-RS2 & 21 & 0.0345 & 0.62  \\
5 & KM-RS1 & 83.3 & 0.0653 & 1.12  \\
6 & KM-RS2 & 125 & 0.0695 & 1.23 \\
\bottomrule
\end{tabular}%
}
\end{table}

\subsection{Comparisons of model predictions}
\begin{figure}
    \centering
    \includegraphics[width=0.4\linewidth]{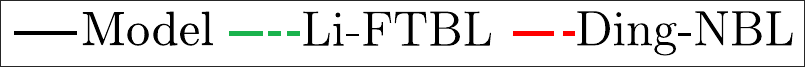}\\
    \vspace{0.3cm}
    \includegraphics[width=0.8\linewidth]{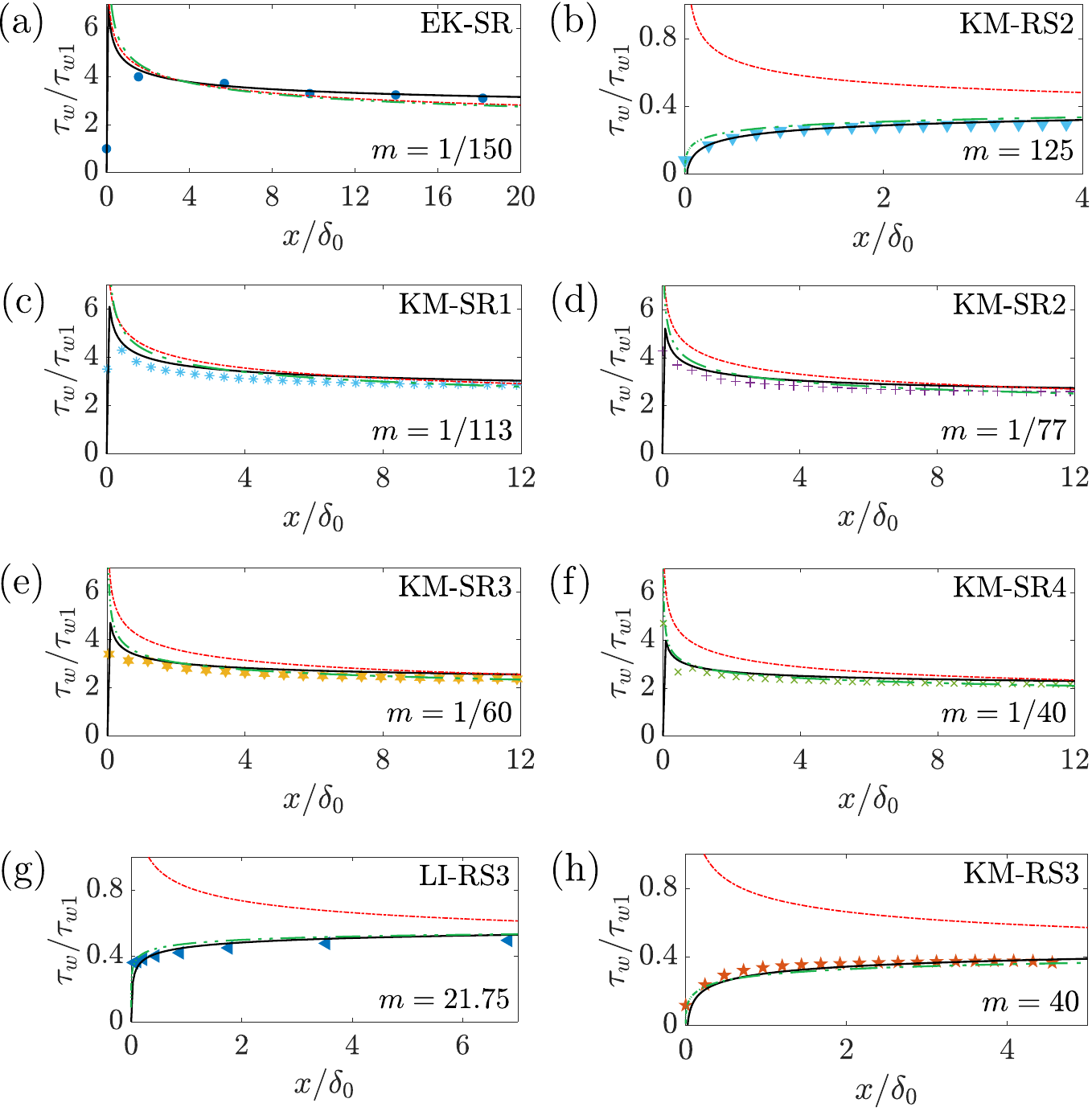}
    \caption{Comparison of the experiments/LES (symbols as in Tables~\ref{tab:cases_train}, \ref{tab:cases_test}) with model predictions (lines) of the streamwise evolution of the normalized wall shear stress, $\tau_w/\tau_{w1}=u_{*2,loc}^2/u_{*1}^2$. Coefficients $\beta$ and $A_1$ are computed using the fits, Eqs.~\eqref{eq:betafit}, \eqref{eq:A1fit}}
    \label{fig:model_tauw}
\end{figure}

\begin{figure}
    \centering
    \subfigure[Model]{\includegraphics[width=0.47\linewidth]{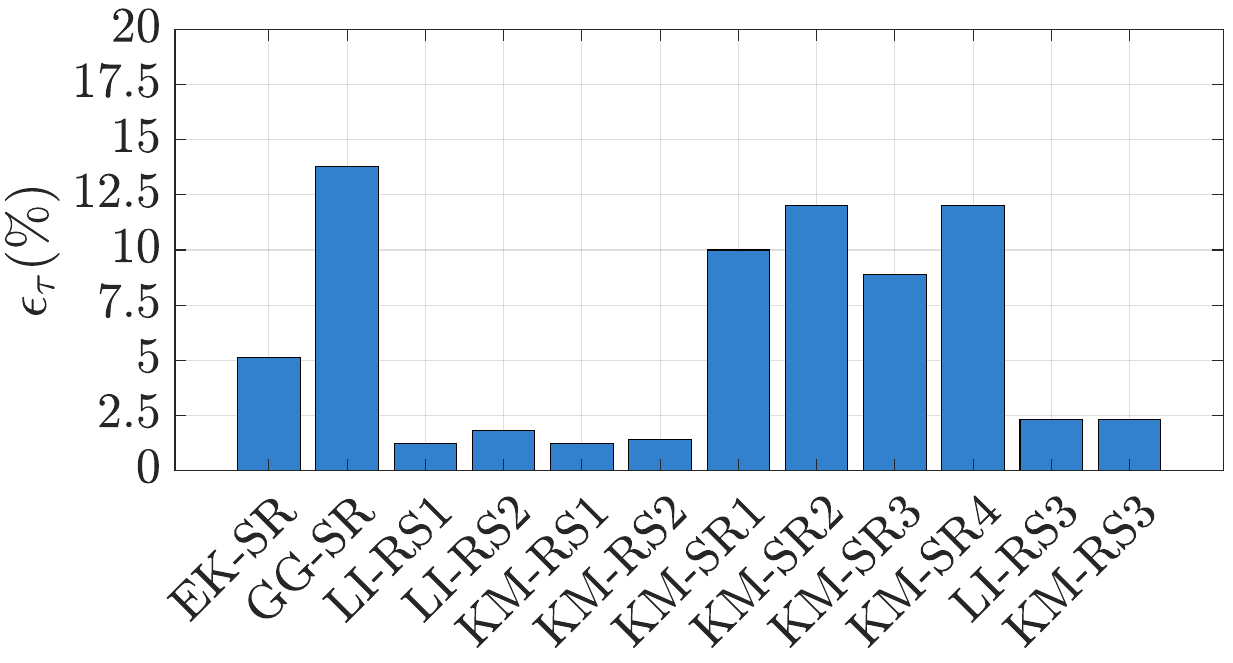}}
    \subfigure[Li-FTBL]{\includegraphics[width=0.47\linewidth]{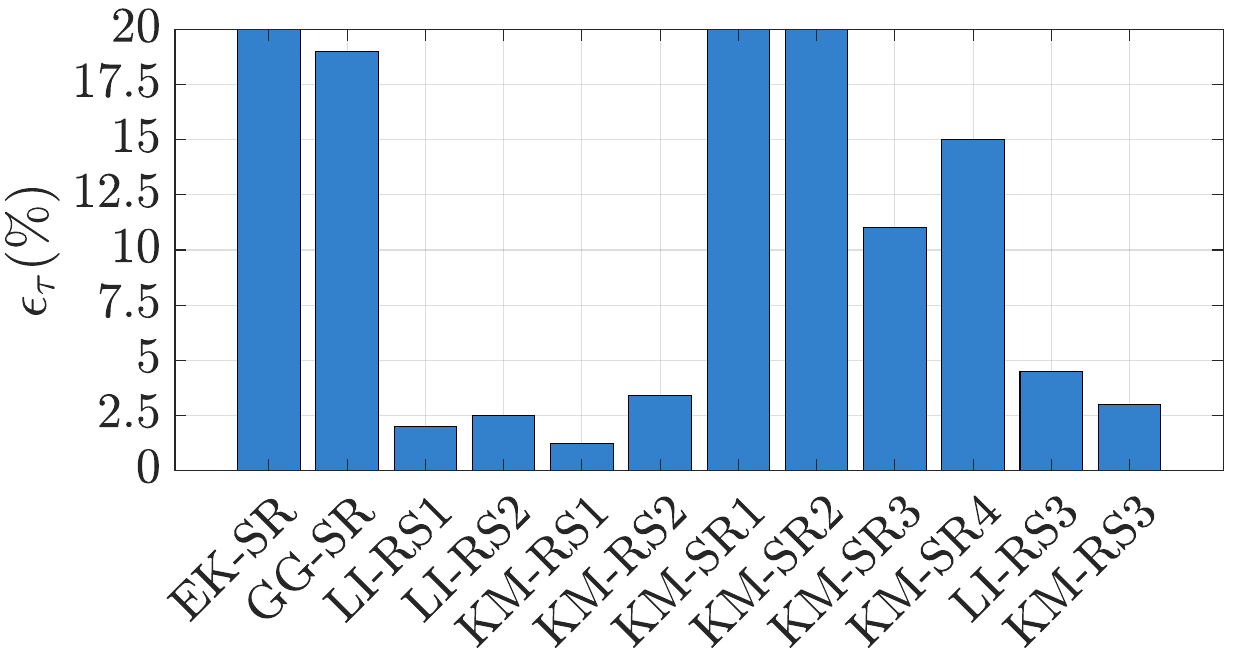}}
    \subfigure[Ding-NBL]{\includegraphics[width=0.47\linewidth]{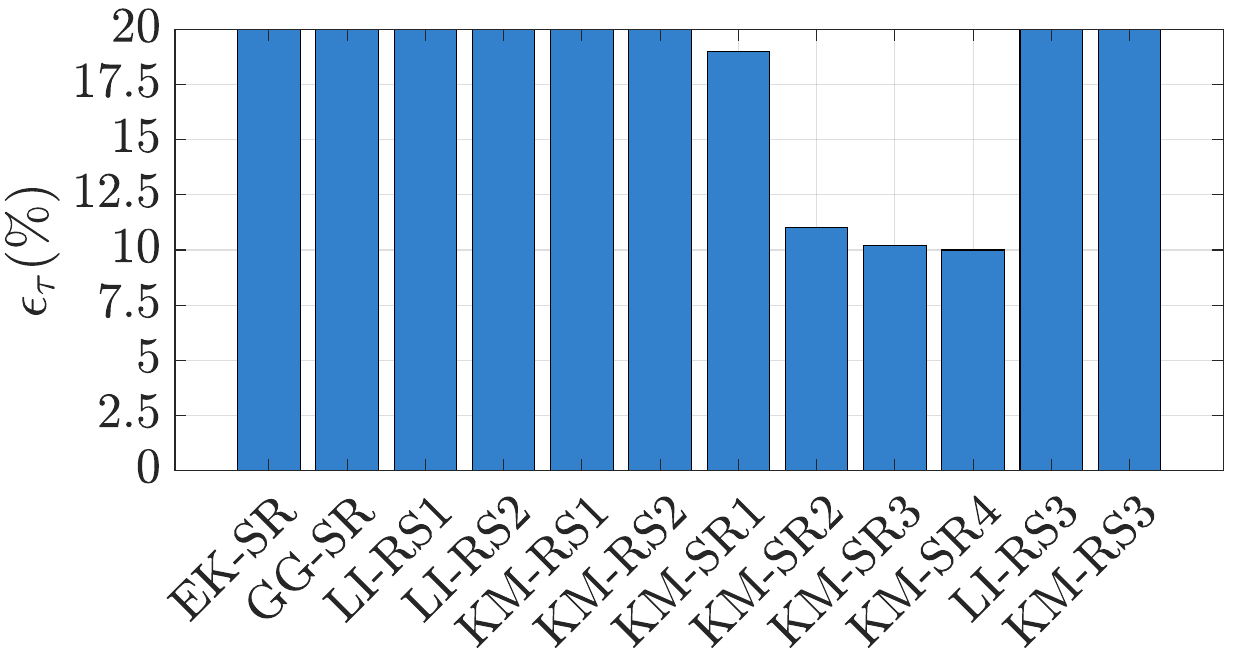}}
    \caption{$L_2$ norms of the difference between the model predictions and experimental/LES data for the wall shear stress downstream of the roughness transition across all the cases listed in Tables~\ref{tab:cases_train} and \ref{tab:cases_test}}
    \label{fig:err_tau}
\end{figure}

\begin{figure}
    \centering
    \includegraphics[width=0.4\linewidth]{figures/tauw_vel_legend.pdf}\\
    \vspace{0.3cm}
    \includegraphics[width=0.9\linewidth]{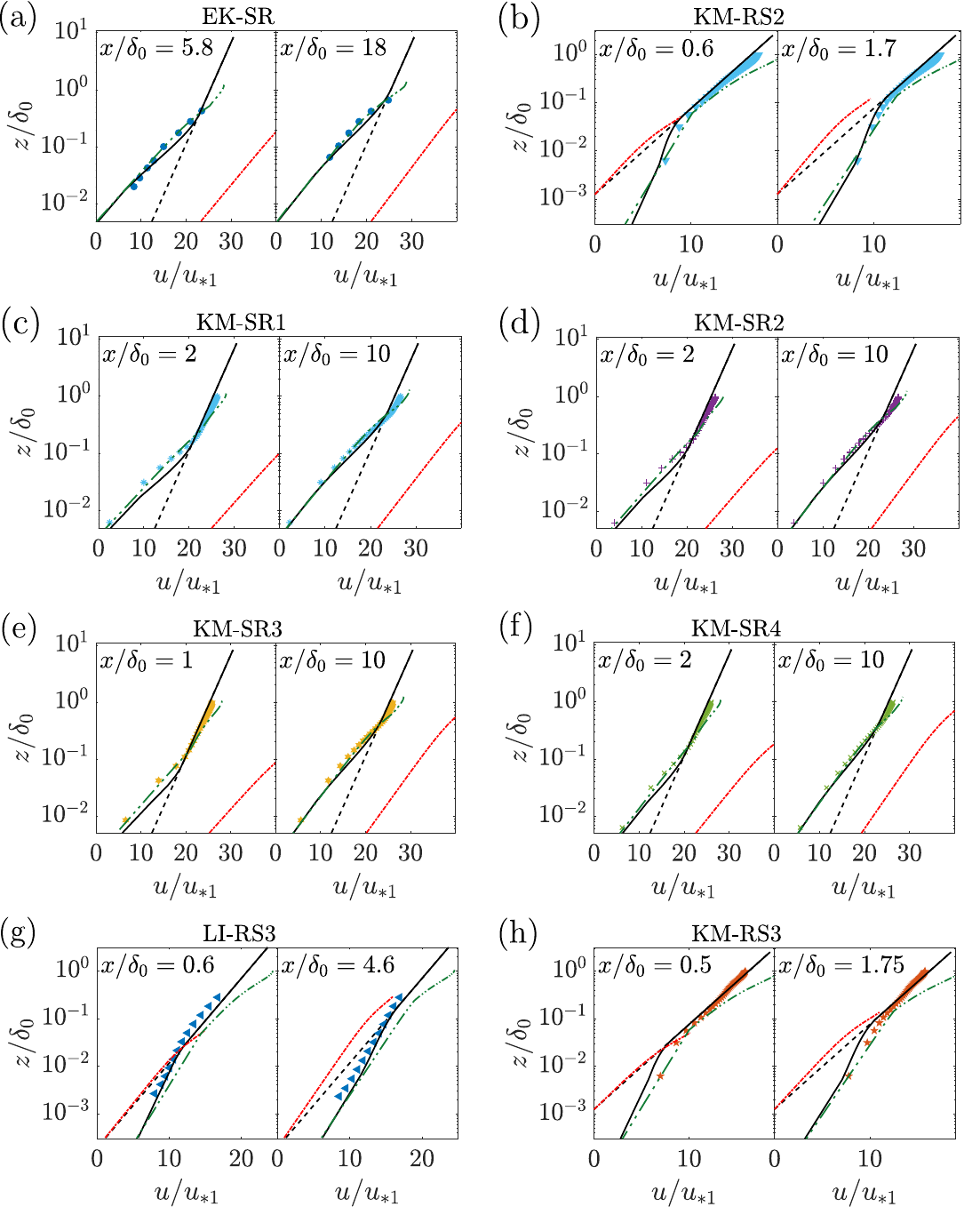}
    \caption{Comparison of the experiments/LES (symbols as in Tables~\ref{tab:cases_train}, \ref{tab:cases_test}) with model predictions (lines) of vertical profiles of the normalized mean streamwise velocity ($\bar{u}/u_{*1}$) at two locations downstream of the roughness transition. Coefficients $\beta$ and $A_1$ are computed using the fits, Eqs.~\eqref{eq:betafit}, \eqref{eq:A1fit}. The black dashed line represents the logarithmic law for the streamwise velocity corresponding to the upstream roughness, Eq.~\eqref{eq:u_eqb_upper}}
    \label{fig:model_u}
\end{figure}

The current model is evaluated against two cases from set A (Table~\ref{tab:cases_train}) and six cases from set B (Table~\ref{tab:cases_test}). The remaining four cases of set A are omitted here for brevity, but are provided in a supplementary document. It should be noted that the set B cases represent completely unseen data for the model since none of these datasets were used to determine the functional dependencies explored in Section~\ref{ssec:ibl_model} or to determine the model coefficients in Section~\ref{sec:coeffs}. Furthermore, for the first 5 cases of Set A, experimental data reported in the literature, as noted in Table~\ref{tab:cases_train}, is used rather than our LES of the corresponding experiments. Thus, these datasets have also not been directly employed in developing the current model, and represent a fair evaluation of the current modelling framework. 

We contrast the predictions of our model with those of three other predictive models: \citet{li2022modelling}, \citet{ding2025ibl} and \citet{savelyev2005internal}, referred to as Li-FTBL, Ding-NBL and ST, respectively. The Li-FTBL and Ding-NBL models are chosen for comparison here because they are the only comprehensive models that predict all three quantities of interest, namely mean velocity, wall shear stress and IBL height. We note that the Ding-NBL model is obtained by setting $L_o=\infty$ in the model of \citet{ding2025ibl} to apply it to neutral conditions. The ST model is not comprehensive since it only predicts the IBL height. However, it is chosen here for comparison since our study presents a direct modification of this model. 

Figure~\ref{fig:model_tauw} compares the model predictions of wall shear stress, $\tau_w/\tau_{w1}=u_{*2,loc}^2/u_{*1}^2$, downstream of the roughness transition. Our proposed model (black solid line) and Li-FTBL (green dashed-dot-dot line) show good agreement with the experiments/LES, fairly capturing the overshoot in $\tau_w$ for the S$\rightarrow$R cases and the undershoot in R$\rightarrow$S cases near $x=0$, followed by an asymptotic approach to a value above or below one at large $x$ values. The Ding-NBL (red dashed-dot line) model is qualitatively correct only for the S$\rightarrow$R cases and fails to capture the undershoot in the wall shear stress in the R$\rightarrow$S cases (Fig.~\ref{fig:model_tauw}(b), \ref{fig:model_tauw}(g), \ref{fig:model_tauw}(h)). This is because the Ding-NBL model is based on empirical relations for $\sigma_w$ and $\bar{u}$ that were obtained using fitting across experiments for only S$\rightarrow$R transition cases \cite{ding2025ibl}. Figure~\ref{fig:err_tau} shows the $L_2$ norms of the errors calculated between the experiments/LES and the model predictions for $\tau_w$ across all twelve cases. The $L_2$ norms across R$\rightarrow$S cases for the present model are less than 2.5\%, whereas the errors are below 15\% for S$\rightarrow$R cases. The Li-FTBL model performs well, with errors below 5\% for R$\rightarrow$S cases but exhibits error values more than 20\% for the S$\rightarrow$R cases. The Ding-NBL performs poorly, with errors exceeding 20\% across most cases, except for four S$\rightarrow$R cases where the errors range from roughly $10\%$ to $19\%$.

The streamwise velocity predictions at two different locations downstream of the roughness transition are presented in Figure~\ref{fig:model_u}. In each panel, the black dashed line represents the logarithmic law of the wall corresponding to the upstream conditions, Eq.~\eqref{eq:u_eqb_upper}. Following the roughness transition, the flow accelerates or decelerates for an R$\rightarrow$S or S$\rightarrow$R transition, respectively, up to the IBL height. Above the IBL height, the downstream velocity profile adheres to the upstream logarithmic profile. Figure~\ref{fig:model_u} shows that this behaviour is captured well by the present model, showing good qualitative and quantitative agreement with the experimental/LES data for all cases at both locations. The Li-FTBL predictions also generally agree well with the experimental/LES data. Below the IBL height, the Li-FTBL predictions are in several cases more accurate than those of our model, but the Li-FTBL predictions deviate from the reference results above $\delta_i$ where they show an acceleration of the flow. This is seen, e.g., in Figures~\ref{fig:model_u}(b,e,h) and can be attributed to the modelling approach in \citet{li2022modelling}. This model was developed assuming a ratio of the outer boundary-layer height to the IBL height, $\delta_c/\delta_i$, of 1.2, which is much smaller than that observed in an ABL flow.  
In contrast to the present model and the Li-FTBL model, the Ding-NBL fails to capture the observed flow behaviour across the cases considered.

\begin{figure}
    \centering
    \includegraphics[width=0.47\linewidth]{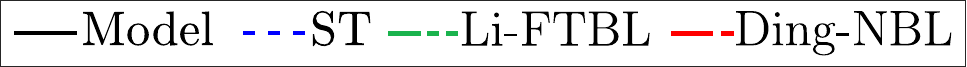}\\
    \vspace{0.3cm}
    \includegraphics[width=0.8\linewidth]{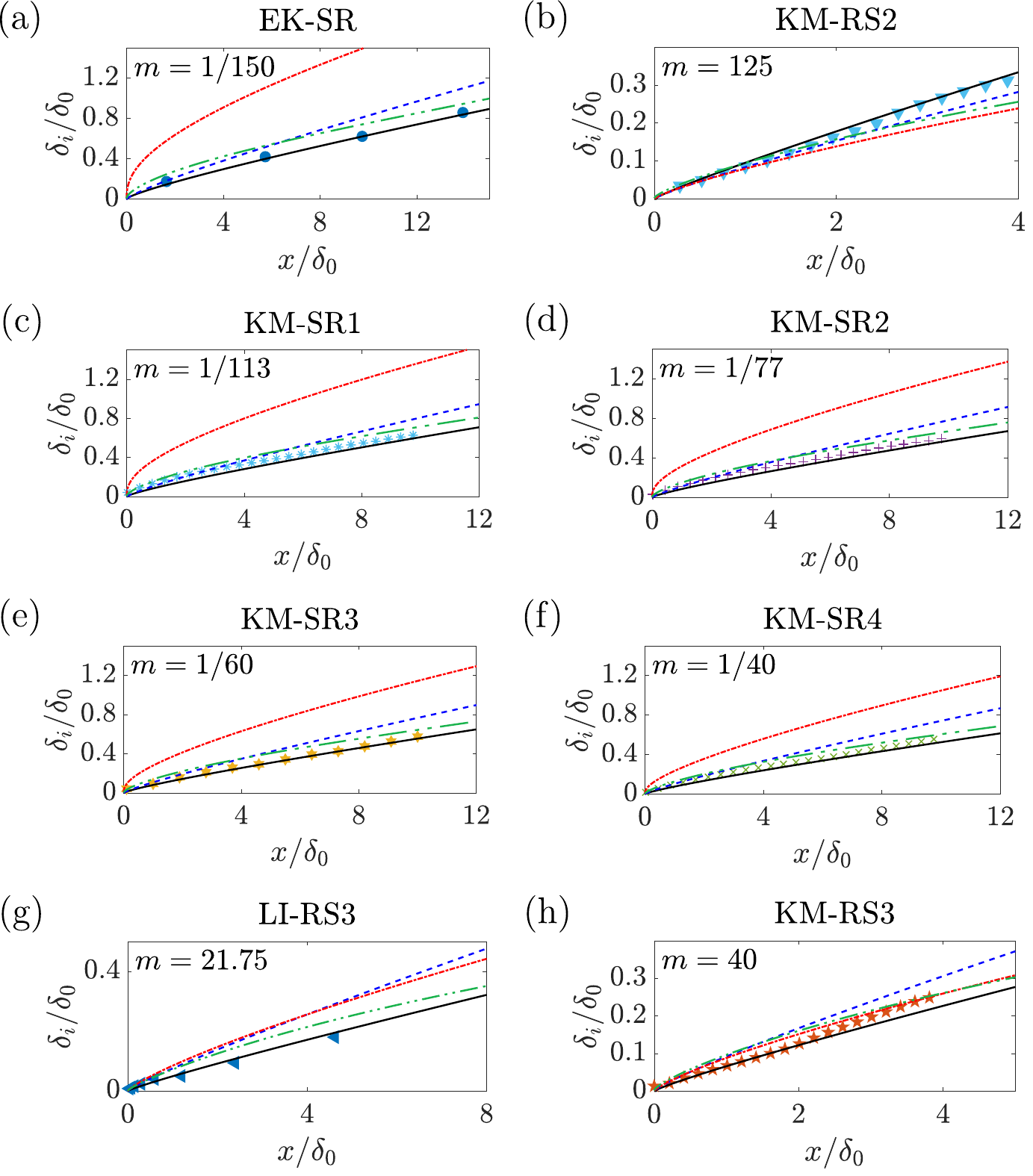}
    \caption{Comparison of the experimental/LES data (symbols as in Tables~\ref{tab:cases_train}, \ref{tab:cases_test}) with model predictions (lines) of the streamwise evolution of the IBL height, $\delta_i$. Coefficients $\beta$ and $A_1$ are computed using the fits, Eqs.~\eqref{eq:betafit}, \eqref{eq:A1fit}}
    \label{fig:deli}
\end{figure}

\begin{figure}
    \centering
    \subfigure[Model]{\includegraphics[width=0.47\linewidth]{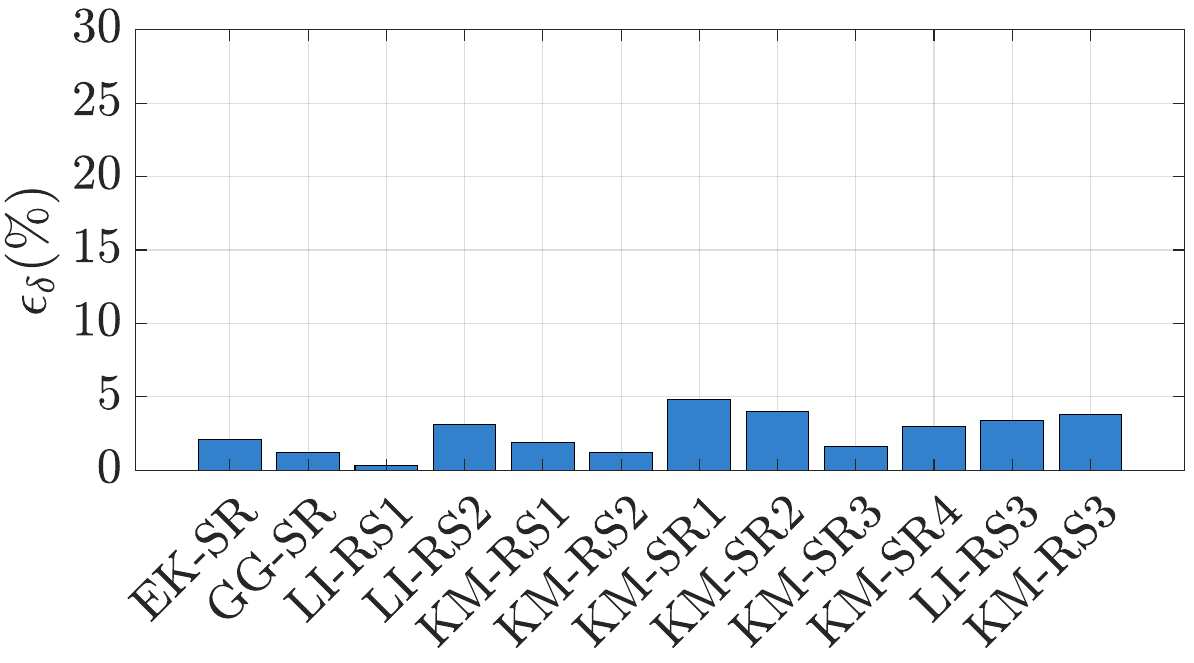}}
    \subfigure[ST]{\includegraphics[width=0.47\linewidth]{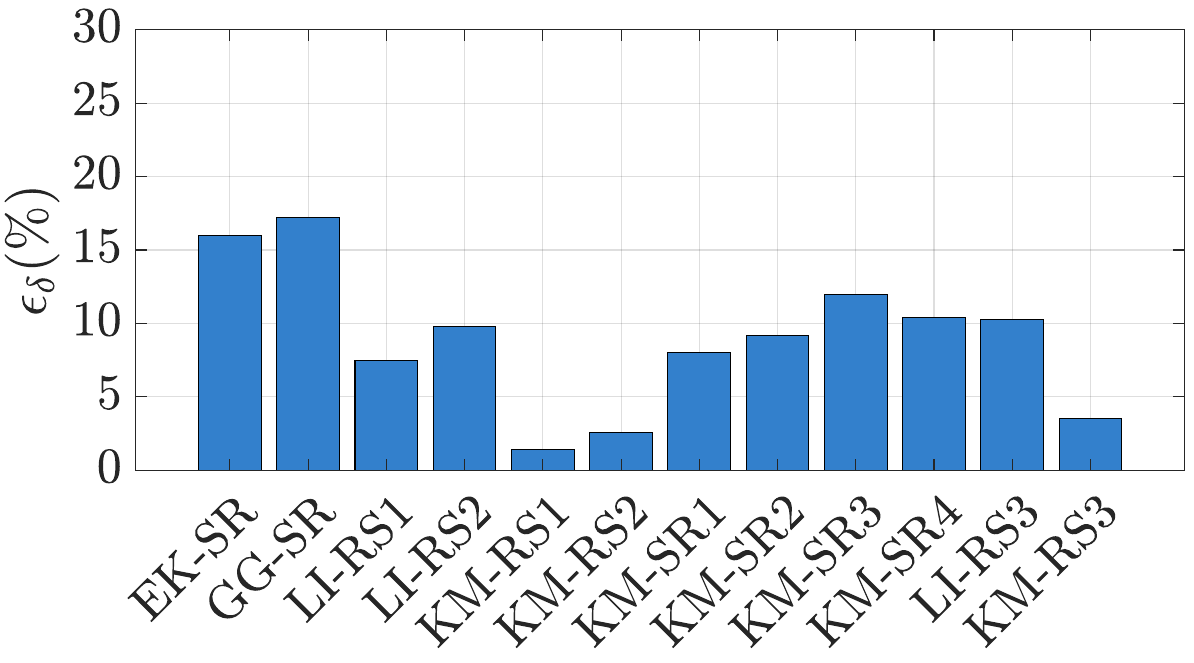}}
    \subfigure[Li-FTBL]{\includegraphics[width=0.47\linewidth]{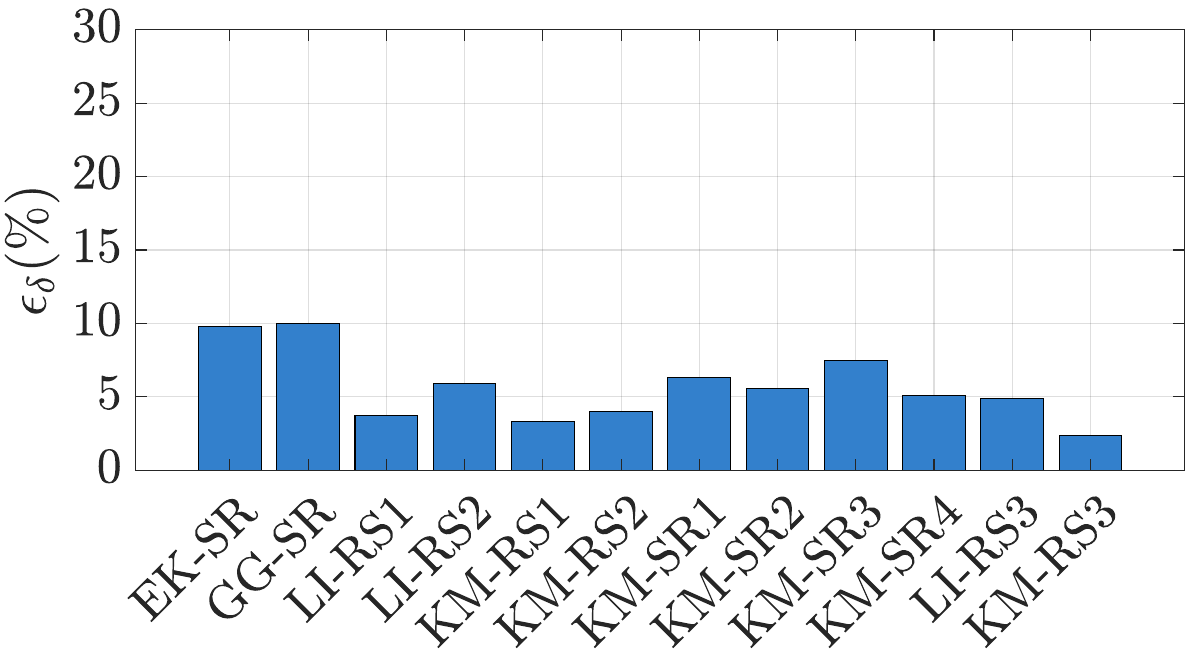}}
    \subfigure[Ding-NBL]{\includegraphics[width=0.47\linewidth]{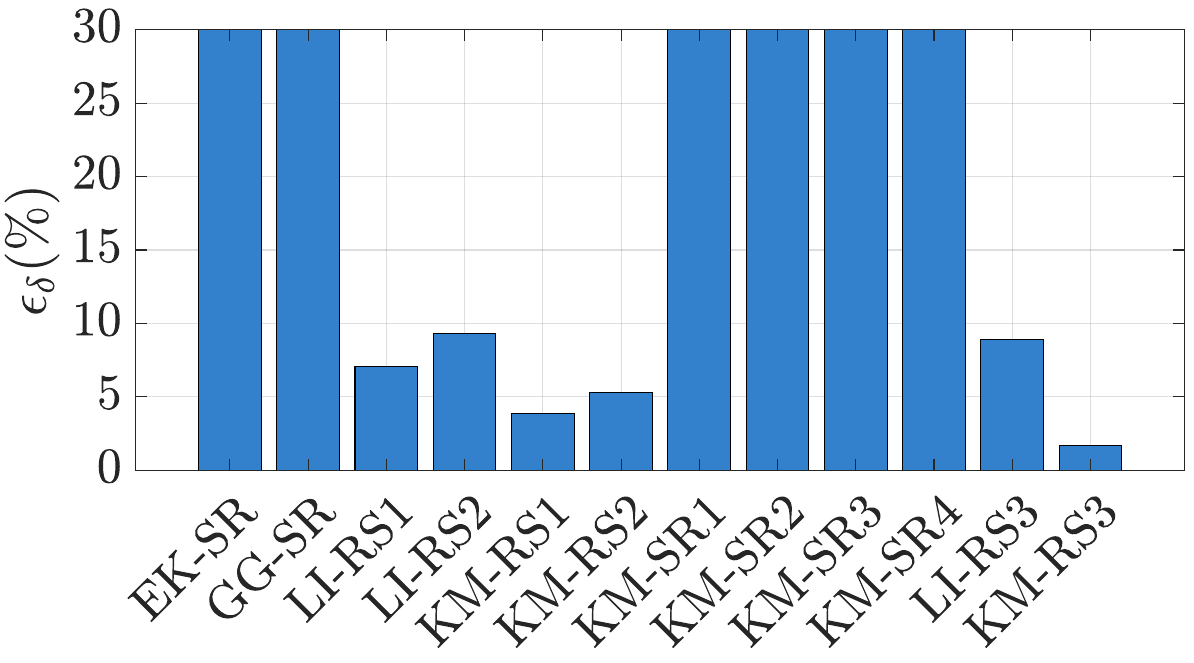}}
    \caption{$L_2$ norms of the difference between the model predictions and experimental/LES data for the IBL height, $\delta_i$, downstream of the roughness transition across all the cases listed in Tables~\ref{tab:cases_train} and \ref{tab:cases_test}}
    \label{fig:err_deli}
\end{figure}

Figure~\ref{fig:deli} compares the IBL height predictions by the ST, Li-FTBl, Ding-NBL and the present models with the experimental/LES data and Figure~\ref{fig:err_deli} shows the corresponding $L_2$ errors. The ST model does not provide a way of predicting the wall shear stress or the mean velocity profile and so was not included in the comparisons shown above. The ST model predictions of the IBL height (blue dashed line) exhibit small error values (less than 5\%) for only three cases (KM-RS1, KM-RS2 and KM-RS3), while substantially over-predicting the $\delta_i$ values in the remaining cases, with $L_2$ error norms exceeding 10\%. Similarly, the Ding-NBL model performs poorly (errors more than $5\%$) across cases except for the same three cases. The Li-FTBL model exhibits fairly good agreement with the reference results, with $L_2$ errors consistently below 10\%. Our present model gives the most accurate predictions, with $L_2$ errors consistently below 5\% across all twelve cases.  

These observations demonstrate that the present formulation is a significant improvement over the model of \citet{savelyev2005internal} and can accurately predict the wall shear stress, mean velocity profiles and IBL height for arbitrary combinations of upstream and downstream roughness values. Furthermore, this model is more generally applicable than the model of \citet{ding2025ibl} and is of comparable or better accuracy than the model of \citet{li2022modelling}.

%% file: files/conclusion.tex
A coupled analytical model is developed to simultaneously predict the downstream friction velocity, the internal boundary-layer (IBL) height and the mean streamwise velocity profile behind an abrupt roughness transition in a neutral ABL. The present formulation significantly modifies the \citet{savelyev2005internal} model by accounting for the local downstream friction velocity in the turbulent diffusion and the mean vertical advection terms, thereby making it physically consistent with the flow adjustment behind a roughness transition. The modelling choices are supported by correlations extracted using LES data covering a range of surface roughness ratios and upstream roughness values. The analytical framework needs prescription of two tunable parameters, namely $\beta$, and $A_1$, which are parameterised as functions of the roughness ratio, $m$. The model is thus fully predictive and generalizes to R$\rightarrow$S and S$\rightarrow$R transitions and a variety of realistic roughness conditions.

Three analytical models, namely \citet{savelyev2005internal} (ST), \citet{li2022modelling} (Li-FTBL) and \citet{ding2025ibl} (Ding-NBL), are evaluated along with the present model against different experimental/LES datasets covering a wide range of roughness ratio values. The present model and the Li-FTBL model exhibit good agreement with the reference experiment/LES, capturing the overshoot/undershoot  (behind a smooth-to-rough/rough-to-smooth transition) in the wall shear stress. They also perform well at predicting the acceleration/deceleration in the mean flow downstream of the roughness transition across all cases. In contrast, Ding-NBL performs poorly at predicting both the wall shear stress and the streamwise velocity. The present model accurately predicts the IBL height for all the experimental and LES cases considered, with $L_2$ errors consistently below 5\%, outperforming the other analytical models, which overpredict the IBL height in most cases.

The current model can serve as a building block in applications such as wind power and air pollution meteorology where effects of surface roughness heterogeneity are important, e.g., coastal effects induced by a shoreline on a near-shore offshore wind farm. The present formulation is restricted to neutral ABL and a single roughness transition. Modifying it to include thermally stratified conditions will be considered in the future. Future work can also extend this model for a stripe-like heterogeneity with successive roughness transitions in the streamwise direction.

%% file: files/app1.tex
This appendix discusses a detailed derivation of the mean streamwise velocity profile in the transition region. Using Eqs.~\eqref{eq:ust_piecewise} and \eqref{eq:nut_trans}, the velocity gradient within the transition region, ($\delta_e \leq z < \delta_i$) is given by
\begin{equation}
    \frac{\partial \bar{u} }{\partial z} = \frac{-\tau}{\nu_t} = -\frac{(u_{*2,\mathrm{loc}}(x)
+\xi\left(u_{*1}-u_{*2,\mathrm{loc}}(x)\right))^2}{-2\beta\kappa(u_{*2,\mathrm{loc}}\delta_e+u_{*1}\delta_i)\xi^2 + [u_{*1}\delta_i(1+2\beta)-u_{*2,\mathrm{loc}}\delta_e(1-2\beta)]\kappa\xi + u_{*2,\mathrm{loc}}\kappa\delta_e}.
\end{equation}
Rearranging the above expression we get 
\begin{equation}
\label{eq:vel_grad_tran}
    \frac{\partial \bar{u}}{\partial z} = \frac{-\tau}{\nu_t} = \frac{(u_{*1}-u_{*2,\mathrm{loc}})^2}{-2\beta\kappa(u_{*2,\mathrm{loc}}\delta_e+u_{*1}\delta_i)} \left[ \frac{\xi^2+p\xi+q}{\xi^2+r\xi+s} \right]
\end{equation}
where,
\begin{equation*}
p = 2\left( \frac{u_{*2,\mathrm{loc}}}
{u_{*1}-u_{*2,\mathrm{loc}}} \right),\;\;\;
q = \left( \frac{u_{*2,\mathrm{loc}}}
{u_{*1}-u_{*2,\mathrm{loc}}} \right)^2 
\end{equation*}
\begin{equation*}
    r = -\frac{(u_{*1}\delta_i(1+2\beta)
-u_{*2,\mathrm{loc}}\delta_e(1-2\beta))}
{2\beta(u_{*1}\delta_i+u_{*2,\mathrm{loc}}\delta_e)},\;\;\;
s = -\frac{u_{*2,\mathrm{loc}}\delta_e}{2\beta(u_{*1}\delta_i+u_{*2,\mathrm{loc}}\delta_e)}.
\end{equation*}
Integrating Eq.~\eqref{eq:vel_grad_tran} gives the streamwise velocity profile given by Eq.~\eqref{eq:u_model}.

For $\beta=0$, the eddy viscosity in the transition region becomes
\begin{equation}
\label{eq:nut_trans_bet0}
    \nu_t^{\mathrm{trans}}(\xi) = (u_{*1}\delta_i-u_{*2,\mathrm{loc}}\delta_e)\kappa\xi+u_{*2,\mathrm{loc}}\kappa\delta_e.
\end{equation}
Using Eq.~\eqref{eq:nut_trans_bet0}, the velocity gradient in the transition region is written as,
\begin{equation}
\label{eq:vel_grad_tran_bet0}
    \frac{\partial \bar{u}}{\partial z} = \frac{-\tau}{\nu_t} = \frac{(u_{*1}-u_{*2,\mathrm{loc}})^2}{\kappa(u_{*1}\delta_i-u_{*2,\mathrm{loc}}\delta_e)} \left[\frac{\xi^2+p\xi+q}{\xi+s^\prime} \right],
\end{equation}
where $p$ and $q$ are defined above and $s^\prime = u_{*2,\mathrm{loc}}\delta_e/(u_{*1}\delta_i-u_{*2,\mathrm{loc}}\delta_e)$. The integrated velocity profile for $\beta=0$ is 
\begin{align}
    \bar{u}(\xi) =\;&\frac{u_{*1}}{\kappa}\ln{\left(\frac{\delta_i}{z_{01}}\right)} + \frac{(u_{*1}-u_{*2,\mathrm{loc}})^2(\delta_i-\delta_e)}{\kappa(u_{*1}\delta_i-u_{*2,\mathrm{loc}}\delta_e)} \Bigg[ \frac{\xi^2-1}{2} + (p-s^\prime)(\xi-1) \nonumber \\[6pt]
    &\qquad + \Bigg\{ \frac{u_{*1}u_{*2,\mathrm{loc}}(\delta_i-\delta_e)}{(u_{*1}-u_{*2,\mathrm{loc}})(u_{*1}\delta_i-u_{*2,\mathrm{loc}}\delta_e)} \Bigg\}^2 \ln \left| \frac{\xi+s^\prime}{1+s^\prime} \right| \Bigg].
\end{align}